\documentclass[aps,showpacs,showkeys,superscriptaddress,preprintnumbers]{revtex4}
\usepackage{amsfonts}
\usepackage{amssymb}
\usepackage{amsmath}
\usepackage{epsfig}
\usepackage{float}
\usepackage{subfig}
\usepackage{array}
\usepackage{epstopdf}
\usepackage{mathrsfs}
\usepackage{multirow}
\usepackage[utf8]{inputenc}
\allowdisplaybreaks
\begin{document}

\title{Fractional Holographic Dark Energy Wormholes: A Comprehensive Geometrical, Physical, and Thermodynamic Investigation}

\author{M. Rizwan}
\email[Email: ]{mrizwan.math@gmail.com}
\affiliation{Institute of Mathematics, University of the Punjab, Lahore-54590, Pakistan}

\author{Z. Yousaf}
\email[Email: ]{zeeshan.math@pu.edu.pk}
\altaffiliation{Corresponding~author}
\affiliation{Institute of Mathematics, University of the Punjab, Lahore-54590, Pakistan}

\keywords{Traversable wormholes; Fractional holographic dark energy; Energy conditions; Wormhole thermodynamics; Non-singular cores.}
\pacs{04.20.Gz; 04.20.Jb; 04.50.Kd; 95.36.+x; 98.80.-k}

\begin{abstract}
The discovery of the accelerated expansion of the cosmos has sparked great interest in studying the nature of the mysterious force behind this effect, often called dark energy. Several theories were proposed to study the nature of dark energy, and holographic dark energy stands out among them due to the relation between dark energy density and the principles of quantum gravity and holography. Recent progress made in fractional cosmology has led to the addition of fractional correction terms to the definition of holographic dark energy. Based on such progress, in this study, we investigate the behavior of fractional holographic dark energy in forming exotic spacetimes, especially traversable wormholes, which represent intriguing solutions of Einstein's field equations connecting distinct regions of spacetime. In the present article, a new class of Morris--Thorne wormhole solutions is obtained under the consideration of fractional holographic dark energy as the source of the gravitational field with a varying redshift function in the context of Einstein gravity. To study the nature of wormholes, their geometry, viability, and thermodynamic behavior, a shape function is obtained, and the wormholes are analyzed in detail via embedding diagrams, throat geometry, active gravitational mass, compactness, exoticity factor, energy conditions, conservation law, volume integral quantifier, Kretschmann invariant, and complexity factor. Moreover, thermodynamic properties of the wormholes are studied via the examination of several parameters, including Hawking temperature, wormhole temperature, entropy, energy, work density, and heat flux, aiming to understand the influence of fractional holographic corrections on the stability and evolution of traversable wormhole structures.

\end{abstract}

\maketitle

\section{Introduction}
The accelerated expansion of the Universe is one of the discoveries that have greatly impacted our understanding of modern cosmology and have posed several questions about the nature of the cosmic element driving this acceleration. Type Ia supernova (SN Ia) measurements, Cosmic Microwave Background (CMB) anisotropy, Baryonic Acoustic Oscillation (BAO), and large-scale structure formation strongly indicate that the Universe is now in an accelerated expansion phase. From the point of view of General Relativity (GR) theory, accelerated expansion is explained by the presence of some unknown element called dark energy (DE), which has negative pressure and accounts for about $70\%$ of the total energy density of the Universe. Despite the outstanding observational success, the physical origin and nature of DE are among the most challenging unresolved problems in modern physics \cite{wald2010general,riess1998observational,supernova1999measurements,aghanim2020planck,spergel2003first,yousaf2025viscous}.

There are several theoretical models for the dynamics of DE, which include the cosmological constant, quintessence, models of scalar fields, theories of modified gravity, and holographic models. Among these models, the Holographic DE (HDE) model has become popular among researchers because it is closely related to quantum gravity concepts. According to the holographic principle, the number of degrees of freedom of any physical system is limited by the surface area of its boundary rather than its volume. In accordance with this concept, Cohen et al. \cite{cohen1999effective} argued that there should be a relationship between the ultraviolet and infrared cutoffs in terms of vacuum energy density. Li \cite{li2004model} extended this model by considering the infrared cutoff to be equal to the future event horizon and gave rise to the HDE Model of DE, capable of explaining the late-time accelerated expansion of the Universe. The HDE model is quite popular because it allows one to consider quantum gravitational effects in cosmological dynamics. The different types of infrared cutoff, such as Hubble horizon, particle horizon, Ricci scalar curvature, and Granda--Oliveros cutoff, were considered to explain the evolution of DE and its observational effects \cite{pavon2005holographic,gao2009holographic,granda2009new}. The standard holographic models usually work for the integer-order evolution of cosmological dynamics and do not take into account some non-locality and memory effects which can occur in such complex gravitational systems. This motivates the exploration of fractional approaches in cosmology.

Fractional calculus, an extension of traditional differentiation and integration into non-integer orders of arbitrary values, has been recognized as a promising tool in modeling memory phenomena, non-normal behavior, and non-local processes in recent years. The use of fractional calculus in cosmology can offer a fresh approach to altering current theories of gravitation and DE. Fractional HDE, as an enhancement of HDE, introduces the concept of the fractional parameter into the traditional holographic model to account for the modified scale factor of energy density. This generalized formulation may provide additional flexibility in describing the evolution of the Universe and investigating alternative gravitational structures.

Wormholes (WHs), on the other hand, constitute one of the most interesting predictions of GR, which describes hypothetical tunnel-like bridges between different parts of spacetime. The study by Morris and Thorne \cite{morris1988wormholes} has offered a comprehensive scheme for creating WHs that can serve as tunnels through space, using the shape and redshift functions to define the characteristics of the WH spacetime. To construct a traversable WH, however, one needs exotic matter that violates the null energy condition (NEC). This necessary condition has driven researchers to seek alternative matter types that may reduce the need for exotic matter.

Recently, the formation of traversable WHs in various gravitational theories has received significant attention. De Falco \textit{et al.}~\cite{de2021reconstructing} constructed WHs in extended gravity theories based on curvature and showed that the inclusion of modified geometric terms can be used to support the existence of traversable WHs with reduced need for exotic matter. In Einstein-Cartan gravity theory, Di Grezia \textit{et al.}~\cite{di2017spin} studied the effect of spin and torsion on the WH geometry and observed that torsion effects help in alleviating the violation of NEC. Nashed~\cite{nashed2005wormhole} found the exact solutions of WHs in teleparallel gravity theory and calculated the gravitational energy corresponding to such solutions. Later on, Nashed~\cite{nashed2009self} constructed the self-dual Lorentzian WHs and studied the energetics of such WHs in the context of teleparallel gravity theory. Recently, Battista \textit{et al.}~\cite{battista2024generalized} added the corrections due to the generalized uncertainty principle to the Rastall-Rainbow gravity and analyzed Casimir-supported WHs, showing that the effects of quantum gravity modify the properties of traversable WHs. Recently, Asad \textit{et al.}~\cite{asad2025traversable} have studied the traversable WHs in $f(R)$ gravity considering the combined effects of global monopole charge and energy conditions, showing that modified gravitational effects can enlarge the physically viable parameter space.

Interactions between dark matter, DE, and WH metrics have paved the way for novel research lines in relativistic astrophysics. Recently, Almatroud \textit{et al.}~\cite{almatroud2025decoupling} used the minimal geometric deformation scheme to construct anisotropic WHs immersed in dark matter halos and showed that gravitational decoupling provides a powerful technique to construct physically viable solutions. Motivated by this work, Alshammari \textit{et al.}~\cite{alshammari2026imprints} studied the observational signatures of HDE and minimally deformed WHs within GR theory and focused on the impact of holographic energy density on WH geometry. The geometric nature of WHs has been analyzed by Alblowy \textit{et al.}~\cite{alblowy2026testing}, and the authors proposed the complexity factor analysis as a suitable method to classify physically plausible traversable WHs. Stable solitonic dark matter WHs within non-minimally $f(Q,T)$ gravity were obtained by Nashed and El Hanafy~\cite{nashed2025stable}, and the importance of nonmetricity in constructing stable WHs was stressed. Yousaf \textit{et al.}~\cite{yousaf2025dark} have studied the traversable WHs from the Einasto dark matter density profile, while Albalahi \textit{et al.}~\cite{albalahi2026constructing} extended such models via non-trivial redshift functions, yielding highly sophisticated geometrical features. In addition, Yousaf \textit{et al.}~\cite{yousaf2025minimally} have constructed minimally complex fuzzy WHs in the framework of Einstein--Cartan theory of gravity, whereas Yousaf \textit{et al.}~\cite{yousaf2026repulsive} have analyzed the influence of the combination of repulsive anisotropy and global monopoles on the structure of traversable WHs in modified gravity. Finally, Khan \textit{et al.}~\cite{khan2026torsionally} also considered the effect of torsional correction in anisotropic dark matter halos via gravitational decoupling, and emphasized the importance of torsion to ensure the physical validity of such gravitational systems.

Motivated by the above and recent developments \cite{de2020general,de2021reconstructing}, in this work we investigate a new class of Morris--Thorne (MT) WH solutions supported by fractional HDE within the framework of GR. Unlike conventional HDE models, the fractional formulation introduces an additional degree of freedom through the fractional parameter, allowing a more generalized description of the energy distribution. We consider a variable redshift function and derive the corresponding shape function from the Einstein field equations by incorporating the fractional holographic energy density as the source of the WH geometry. The effects of fractional corrections on the WH structure are explored in detail.

The fundamental purpose of this research is to conduct an extensive analysis of the geometric, physical, and thermodynamic nature of WHs sustained by fractonal HDE. We structure this article in the following way:
\begin{enumerate}
    \item The description of fractional HDE and WH field equations are provided in Section II.
    \item To understand the geometric structure of the models, the nature of the shape function, throat geometry, and embedded surfaces are analyzed in Section III.
    \item The thermodynamics is developed by investigating Hawking, WH temperatures, entropy, energy, work density, and heat flux of the solution in Section IV.
    \item In Section V, the physical nature of the models is explored by examining energy conditions, the equation-of-state (EoS) parameters, compactness, active gravitational mass, and the exoticity parameter, the conservation equations, and the volume integral quantifiers.
    \item Finally, in Section VI, the nature of regularity and structural stability has been investigated via Kretschmann and complexity factors, and Section VII provides the concluding arguments.
\end{enumerate}
 The results obtained in this research provide new insight into the nature of the role of fractional HDE in forming traversable WH geometry and developing a link between quantum-inspired DE models, the geometry of spacetime, and gravitational thermodynamics.

\section{Fractional Holographic Dark Energy and Wormhole Field Equations}

The holographic principle, motivated by black hole thermodynamics and quantum gravity, states that the number of degrees of freedom of a physical system is determined by its boundary area rather than its volume~\cite{hooft1993dimensional,susskind1995world}. Based on this idea, Cohen \textit{et al.}~\cite{cohen1999effective} established a relation between the ultraviolet and infrared cutoffs of an effective field theory, while Li~\cite{li2004model} proposed the HDE model with energy density
\begin{equation}
\rho_{\rm HDE}=3d^{2}M_{p}^{2}L^{-2},
\end{equation}
where $L$ is the infrared cut-off, and $d$ is a dimensionless parameter. Due to the solid theoretical basis and the concurrence with cosmological observations, the HDE model has emerged as one of the most popular candidates for DE~\cite{wang2016dark,li2011dark}.

Recently, Trivedi \textit{et al.}~\cite{trivedi2024fractional} have extended the conventional HDE model by means of fractional calculus, thus formulating the fractional HDE model. The associated density function is defined as follows:
\begin{equation}
\boxed{\rho_{\rm FHDE}=3d^{2}L^{\frac{2-3\gamma}{\gamma}}}.
\label{FHDE}
\end{equation}
Here, $\gamma$ is the fractional parameter, whereas the usual or conventional HDE density formula is recovered when $\gamma=2$. This extra fractional degree of freedom is naturally capable of considering nonlocal interactions and adding some complexity to the underlying cosmology. In addition, Bidlan \textit{et al.}~\cite{bidlan2025reconstructing} have shown that fractional HDE can be reproduced via scalar and gauge field models, implying its viability as a generalized DE model. With this motivation, we use the fractional HDE density as a matter source to construct MT WH solutions in GR and investigate their geometrical and thermodynamic properties.

We analyze the spacetime of the static and spherically symmetric MT WH given by the metric \cite{morris1988wormholes,morris1988wormholess}
\begin{equation}
\boxed{ds^{2}=-e^{2\phi(r)}dt^{2}
+\frac{dr^{2}}{1-\frac{\Psi(r)}{r}}
+r^{2}\left(d\theta^{2}+\sin^{2}\theta\,d\phi^{2}\right).}
\label{metric}
\end{equation}
Here $\phi(r)$ is the redshift function, and $\Psi(r)$ is the shape function. The redshift function must be finite everywhere in the spacetime to avoid having any horizons and WH traversability, whereas the shape function gives the spatial geometry of the WH. To meet the requirements of the throat of the WH $r=r_{0}$, the following conditions should be met:
\begin{equation}
\Psi(r_{0})=r_{0},
\end{equation}
with the flare-out condition
\begin{equation}
\Psi'(r_{0})<1,
\end{equation}
while asymptotic flatness requires
\begin{equation}
\lim_{r\rightarrow\infty}\frac{\Psi(r)}{r}=0.
\end{equation}

The nature of the material threading the WH is taken to be anisotropic and can be expressed by the energy-momentum tensor
\begin{equation}
\boxed{T_{\beta\zeta}
=(\mu+p_{tn})U_{\beta}U_{\zeta}
+p_{tn}g_{\beta\zeta}
+\left(p_{rd}-p_{tn}\right)\chi_{\beta}\chi_{\zeta},}
\label{EMTensor}
\end{equation}
where $\rho$ denotes the energy density, $p_{rd}$ is the radial pressure, $p_{tn}$ is the tangential pressure, $U^{\beta}$ is the four-velocity with the property $U^{\beta}U_{\beta}=-1$, and $\chi^{\beta}$ is the spacelike radial vector of unit magnitude that satisfies $\chi^{\beta}\chi_{\beta}=1$ and $U^{\beta}\chi_{\beta}=0$.

Throughout this work, we use the geometrized units, $G=c=1$, so that the Einstein field equations can be expressed as
\begin{equation}
\boxed{G_{\beta\zeta}=8\pi T_{\beta\zeta}.}
\label{EFE}
\end{equation}
Plugging the metric (\ref{metric}) into Eq.~(\ref{EFE}), one gets the following field equations independently:
\begin{equation}
8\pi\mu(r)=\frac{\Psi'(r)}{r^{2}},
\label{rho}
\end{equation}

\begin{equation}
8\pi p_{rd}(r)=
-\frac{\Psi(r)}{r^{3}}
+2\left(1-\frac{\Psi(r)}{r}\right)\frac{\phi'(r)}{r},
\label{pr}
\end{equation}

\begin{align}
8\pi p_{tn}(r)=&
\left(1-\frac{\Psi(r)}{r}\right)
\Bigg[
\phi''+\left(\phi'\right)^{2}
+\frac{\phi'}{r}
\nonumber\\
&
-\frac{b'r-b}{2r(r-b)}\Phi'
-\frac{b'r-b}{2r^{2}(r-b)}
\Bigg].
\label{pt}
\end{align}

The conservation equation,
$\nabla_{\mu}T^{\mu}_{\ \nu}=0$, yields

\begin{equation}
p_{rd}'+\phi'(\mu+p_{rd})
+\frac{2}{r}(p_{rd}-p_{tn})=0.
\label{conservation}
\end{equation}
This is the expression that corresponds to the condition of hydrostatic equilibrium for anisotropic matter configurations.

For the construction of a physically realistic traversable WH, a finite and non-constant redshift function rather than the commonly considered constant redshift function is used. The redshift function, which takes into account tidal gravity, gives a more general representation of the WH spacetime without event horizon formation. In our current study, we consider the redshift function to be in the form
\begin{equation}
\phi(r)=-\frac{0.1}{r},
\end{equation}
which is finite for all $r\geq r_0$ such that
\begin{equation}
\lim_{r\rightarrow\infty}\phi(r)=0,
\end{equation}
which ensures asymptotic flatness. For this case, the metric potential is
\begin{equation}
e^{2\phi(r)}=\exp\left(-\frac{0.2}{r}\right),
\end{equation}
which is positive everywhere in the spacetime, indicating that there are no horizons. This choice keeps the analytical simplicity but at the same time allows us to explore the real effects of tidal gravity on the properties of the fractional HDE WHs constructed using the minimal geometric deformation method.

In this analysis, we make use of the density of fractional HDE as the source term, where
\begin{equation}
\mu(r)=\rho_{\rm FHDE},
\end{equation}
and determine the shape function from the above equation assuming a variable redshift function. Hereafter, the geometry of the WH solution is studied using the flare-out condition, embedding procedure, energy conditions, gravitational mass, compactness, exoticity parameter, volume integral quantifier, curvature scalar, complexity factor, and thermodynamics.

On substitution of the density of the fractional HDE, in the $(t,t)$ component of the Einstein field equations, one gets
\begin{equation}
\frac{\Psi'(r)}{8\pi r^{2}}=3d^{2}r^{-3+\frac{2}{\gamma}},
\end{equation}
which can be expressed as
\begin{equation}
\Psi'(r)=24\pi d^{2}r^{-1+\frac{2}{\gamma}}.
\end{equation}
Integration of the above equation with respect to the radial variable gives
\begin{equation}
\Psi(r)=12\pi d^{2}\gamma\,r^{\frac{2}{\gamma}}+h,
\end{equation}
where $h$ is the constant of integration. From the WH throat condition,
\begin{equation}
\Psi(r_{0})=r_{0},
\end{equation}
the constant of integration is obtained as
\begin{equation}
h=r_{0}-12\pi d^{2}\gamma\,r_{0}^{\frac{2}{\gamma}}.
\end{equation}
Hence, the resulting shape function takes the form
\begin{equation}
\boxed{
\Psi(r)=r_{0}+12\pi d^{2}\gamma
\left(r^{\frac{2}{\gamma}}-r_{0}^{\frac{2}{\gamma}}\right).
}
\end{equation}
The derived shape function satisfies the throat condition $\Psi(r_{0})=r_{0}$ exactly. Also, the geometry is clearly controlled by the fractional parameter $\gamma$, implying that the fractional nature of the HDE has a great impact on the WH geometry.

\section{Geometrical Analysis of the Fractional Holographic Dark Energy Wormhole}

After finding the expression for the shape function from the density of HFR, we proceed with the investigation of geometrical characteristics of the derived WH solution. Specifically, we discuss the behavior of the shape function, check the flare-out condition, which is needed for traversability of the WH, and draw the relevant embedding diagrams. In all of this section, the numerical plots will be plotted for
\(
d=0.05,~ r_{0}=1,
\)
with varying fractional parameter \(\gamma=1.25,\;1.5,\;2.,\;2.5,\;3.\)


\subsection{Shape Function Analysis}

\begin{figure}[H]
\centering
\subfloat[]{{\includegraphics[height=2 in, width=3 in]{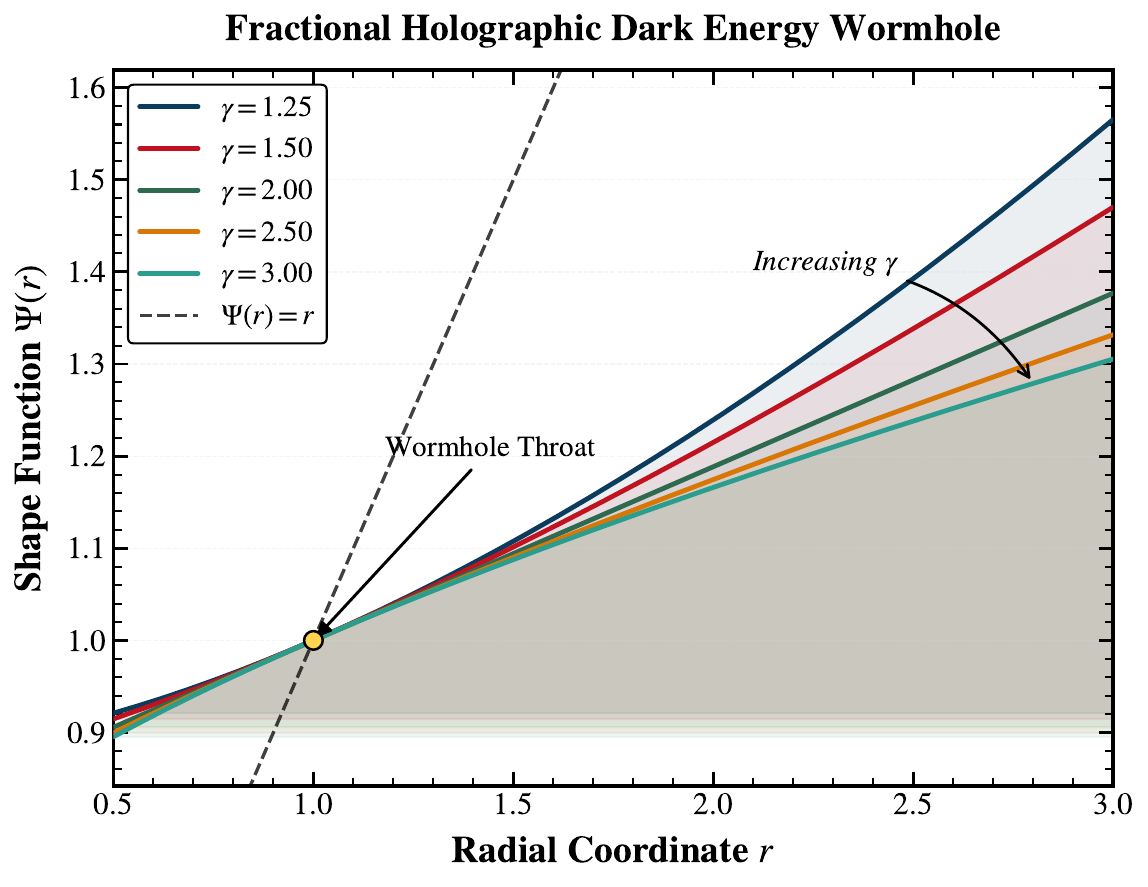}}}
\qquad
\subfloat[]{{\includegraphics[height=2 in, width=3 in]{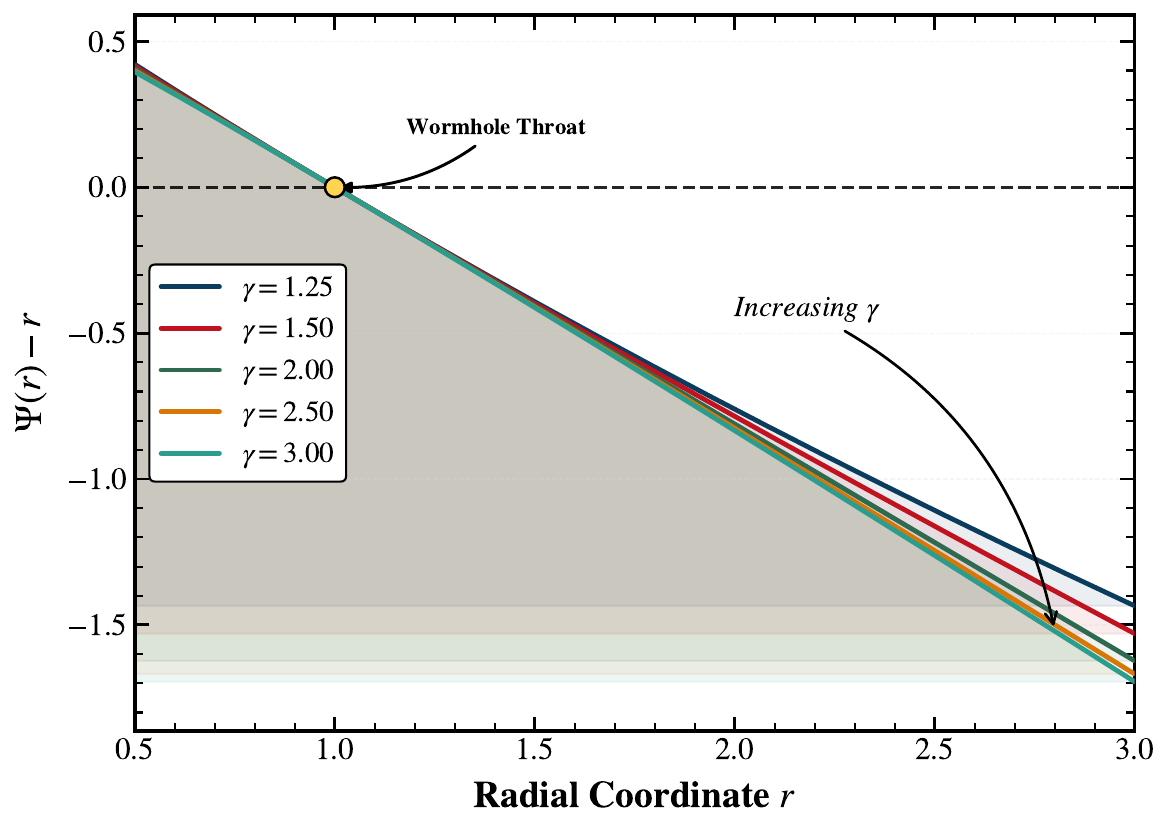}}}
\qquad
\subfloat[]{{\includegraphics[height=2 in, width=3 in]{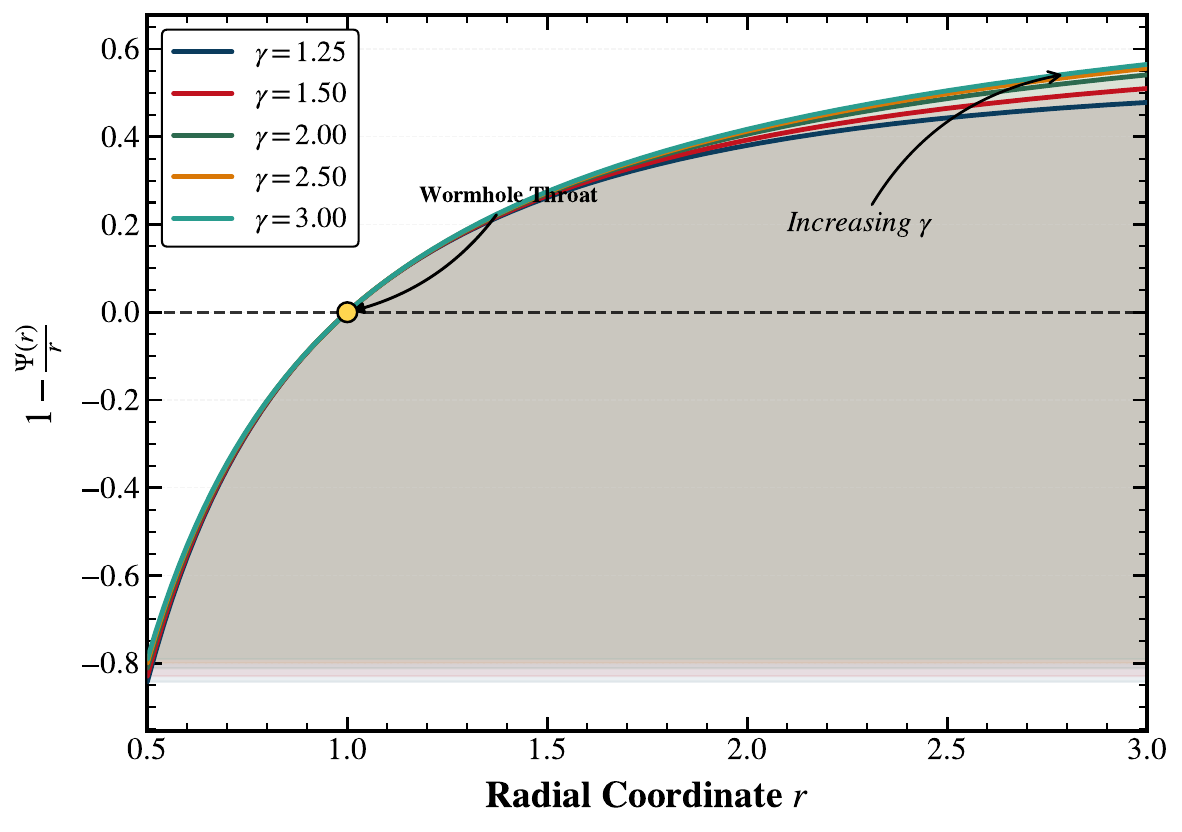}}}
\qquad
\subfloat[]{{\includegraphics[height=2 in, width=3 in]{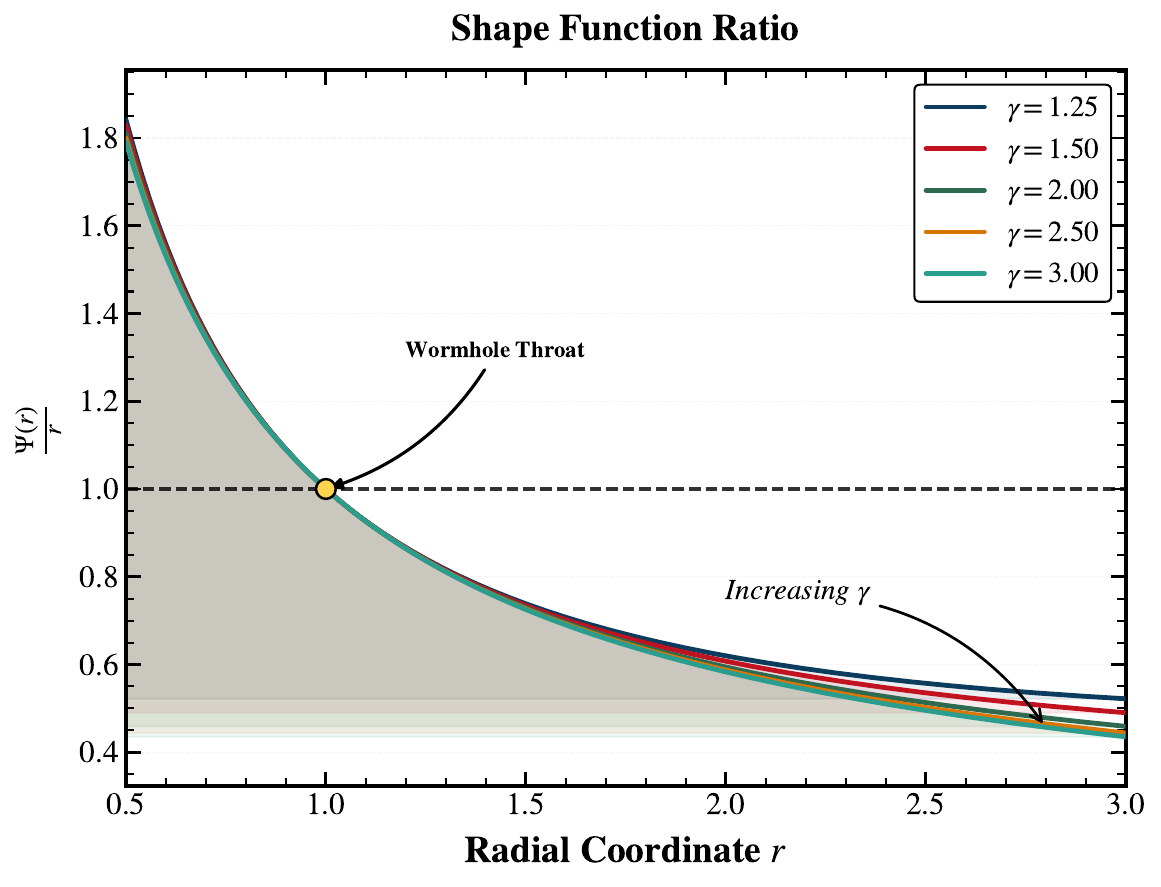}}}
\caption {Geometrical behavior of the fractional HDE WH for different values of the fractional parameter $\gamma$. Panels (a)--(d) depict the shape function $\Psi(r)$, $\Psi(r)-r$, $1-\Psi(r)/r$, and $\Psi(r)/r$, respectively, confirming that all traversability conditions are satisfied, while the fractional parameter $\gamma$ continuously regulates the WH geometry.}\label{1f}
\end{figure}

Geometrical features of the derived fractional HDE WH are demonstrated in Fig. \textbf{\ref{1f}}. The four plots give a complete description of the shape function and check all basic geometrical conditions for WH existence.

In Fig. \textbf{\ref{1f}}(a), the evolution of the shape function $\Psi(r)$ is shown as a function of the fractional parameter $\gamma$. It can be seen that the shape function is an increasing function of the radial coordinate and fulfills the condition
\begin{equation}
\Psi(r_0)=r_0,
\end{equation}
which is shown as an intersection point of the curves at $r_0=1$. As the value of $\gamma$ grows, the increase of $\Psi(r)$ becomes smaller, meaning that larger values of the fractional parameter slow down the expansion of the WH geometry.

In order to confirm that the above-mentioned conditions are satisfied, Fig. \textbf{\ref{1f}}(b) demonstrates the behavior of $\Psi(r)-r$. One can observe that the curves cross the horizontal axis precisely at the WH throat, meaning that $\Psi(r_0)=r_0$. In addition, for $r>r_0$, the expression $\Psi(r)-r$ remains negative, meaning that
\begin{equation}
\Psi(r)<r.
\end{equation}
This means that the coefficient
\begin{equation}
1-\frac{\Psi(r)}{r}
\end{equation}
stays positive everywhere outside the throat. Thus, there is no event horizon, and the obtained solution corresponds to a traversable WH.

The quantity
\begin{equation}
1-\frac{\Psi(r)}{r},
\end{equation}
depicted in Fig. \textbf{\ref{1f}}(c), is positive for any allowed values of $r$. It starts from zero at the throat and monotonically grows up to unity. This means that the spacetime becomes asymptotically flat, while maintaining the regularity of the WH geometry.

Lastly, Fig. \textbf{\ref{1f}}(d) presents the quantity
\begin{equation}
\frac{\Psi(r)}{r},
\end{equation}
which is equal to one at the throat and monotonically decreases when $r$ grows. It is less than one for any value of $\gamma$ beyond the throat, meeting one of the most important MT conditions on the existence of traversable WHs. Moreover, an increase in the value of the fractional parameter causes the slower decrease of $\Psi(r)/r$, meaning that the fractional effects affect the global geometry without violating any necessary conditions on it.

All the behavior mentioned above means that the obtained shape function is geometrically well-defined and meets all the necessary conditions for a physically reasonable fractional HDE WH.


\subsection{Flare-Out Condition}

One of the most fundamental geometrical requirements for a traversable WH is the flare-out condition. At the throat it is expressed as
\begin{equation}
\Psi'(r_{0})<1,
\end{equation}
which guarantees that the spatial hypersurface opens outward rather than pinching off.
\begin{figure}[H]
\centering
\subfloat[]{{\includegraphics[height=2 in, width=3 in]{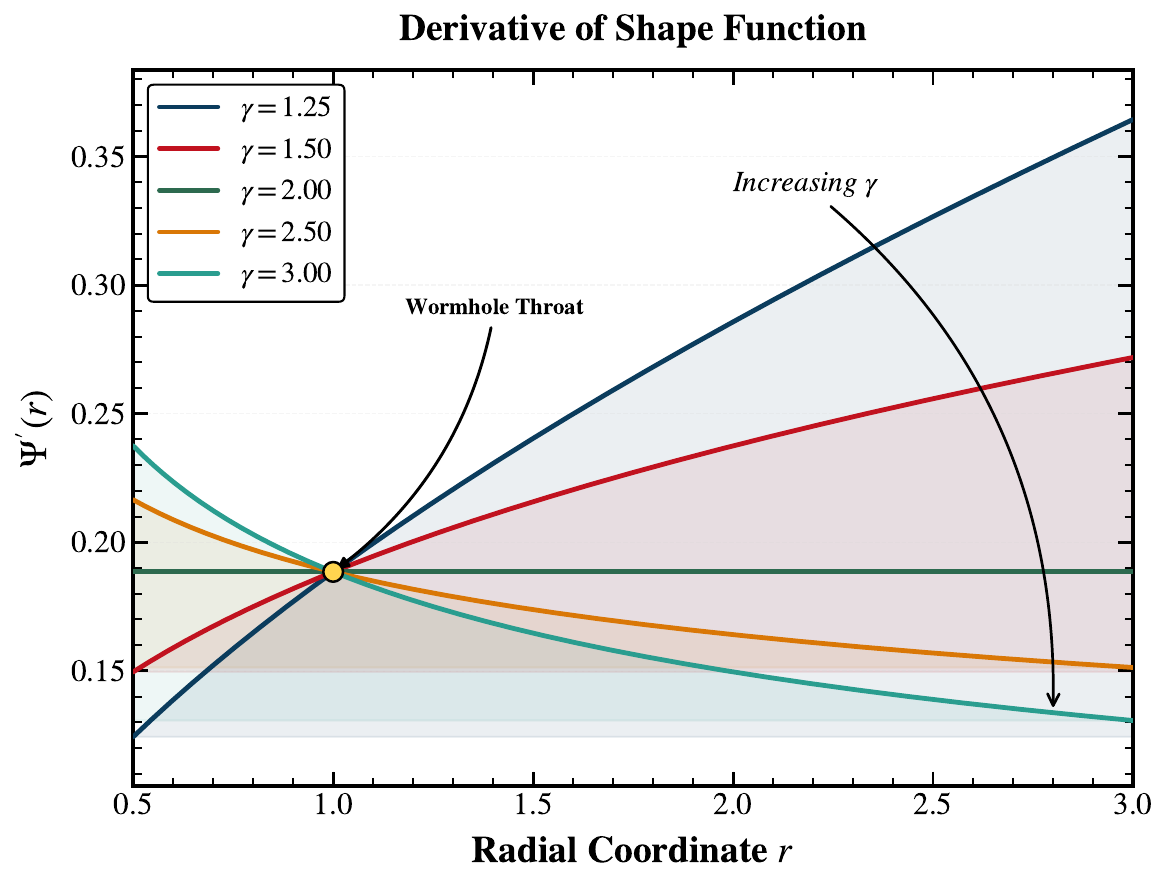}}}
\qquad
\subfloat[]{{\includegraphics[height=2 in, width=3 in]{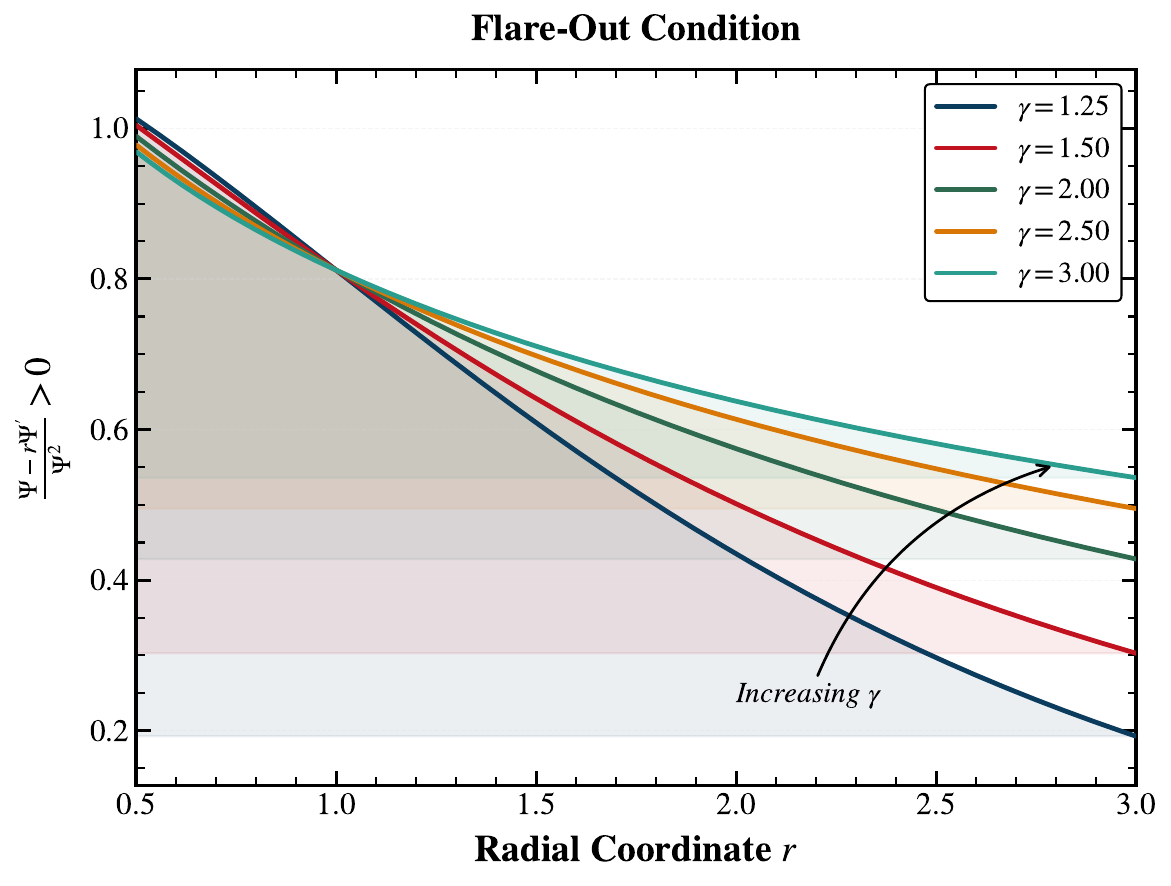}}}
\caption {Graphical depiction of satisfaction of flaring out condition for $d=0.05$, $r_0=1$, and different fractional parameter $\gamma$.}\label{2f}
\end{figure}
The derivative of the obtained shape function is shown in Fig. \textbf{\ref{2f}}(a). Although the derivative exhibits different radial behaviors for different values of the fractional parameter, all curves intersect near the throat with values considerably smaller than unity. Hence the flare-out requirement is fulfilled for every considered model.

The flare-out function,
\begin{equation}
\frac{\Psi-r\Psi'}{\Psi^{2}},
\end{equation}
is plotted in Fig. \textbf{\ref{2f}}(b). It remains strictly positive throughout the physical domain,
\[
\frac{\Psi-r\Psi'}{\Psi^{2}}>0,
\]
thereby providing an independent confirmation that the WH throat is properly flared outward.

An interesting feature of the solutions is the influence of the fractional parameter. Increasing $\gamma$ decreases the slope of the shape function while simultaneously increasing the flare-out function. Consequently, larger values of $\gamma$ correspond to a more gradually flaring throat, although all considered configurations satisfy the MT traversability conditions.


\subsection{Embedding Diagrams}

The embedding plot gives a straightforward geometrical illustration of the spatial geometry of the WH. For an equatorial section, i.e.,
\[
t=\text{constant},\qquad \theta=\frac{\pi}{2},
\]
the embedded surface obeys the following relation \cite{morris1988wormholes,morris1988wormholess,ellis1973ether}:
\begin{equation}
\frac{d\mathcal{Z}}{dr}
=
\pm
\left(
\frac{r}{\Psi(r)}-1
\right)^{-1/2}.
\end{equation}
Here the plus and minus signs correspond to the upper and lower universes that are connected by the WH throat.
\begin{figure}[H]
\centering
\subfloat[]{{\includegraphics[height=2 in, width=3 in]{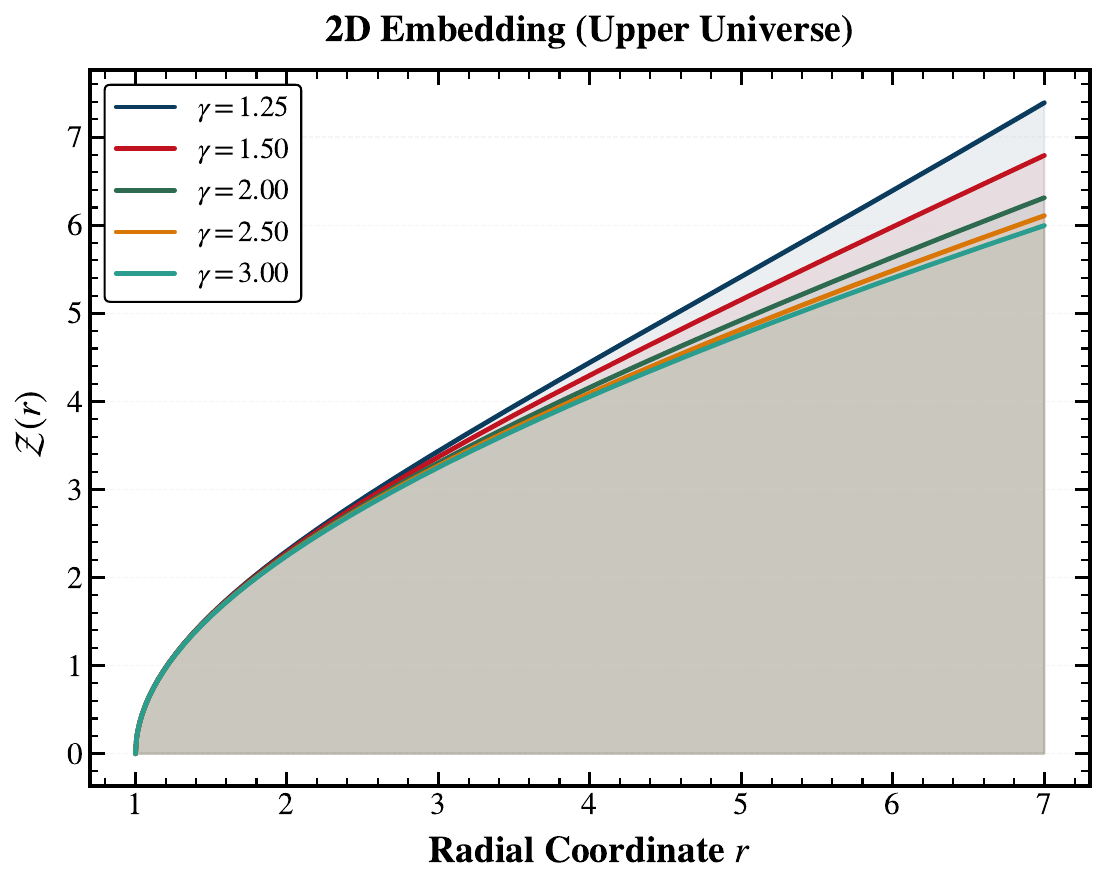}}}
\qquad
\subfloat[]{{\includegraphics[height=2 in, width=3 in]{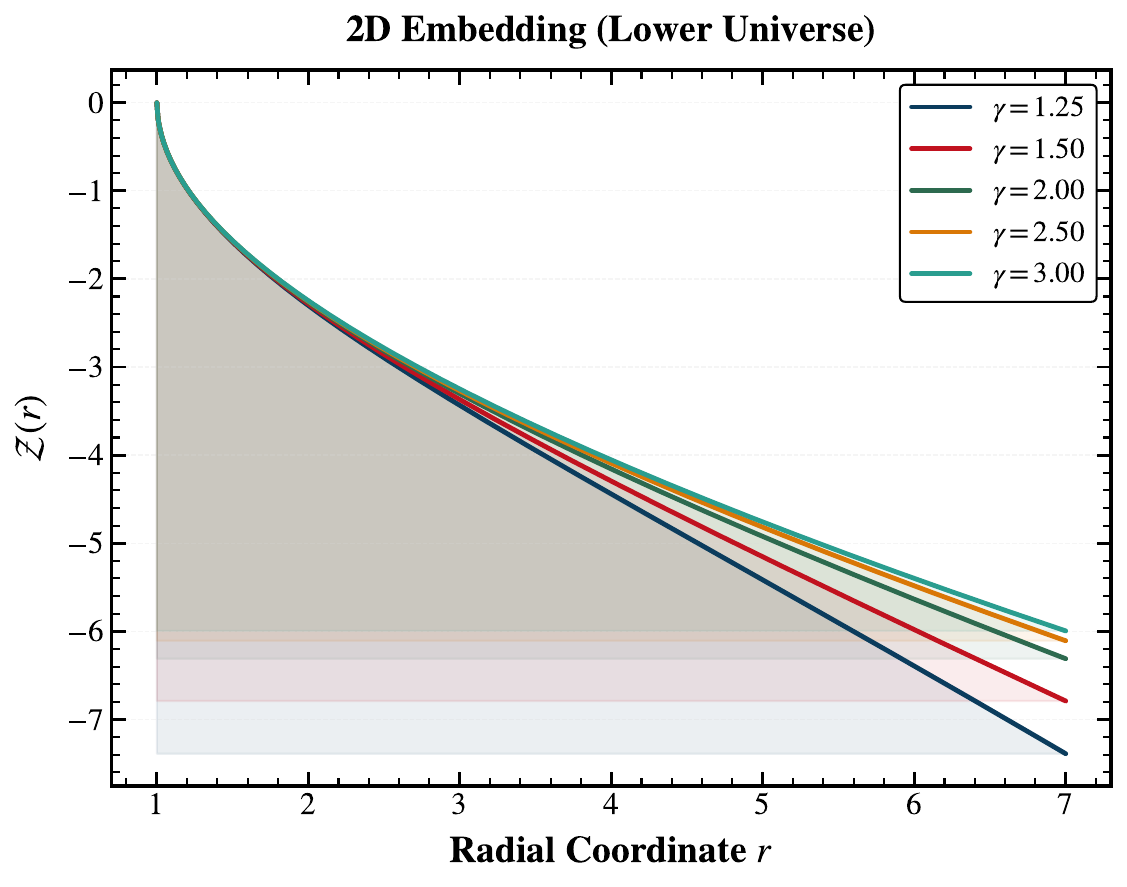}}}
\qquad
\subfloat[]{{\includegraphics[height=2 in, width=3 in]{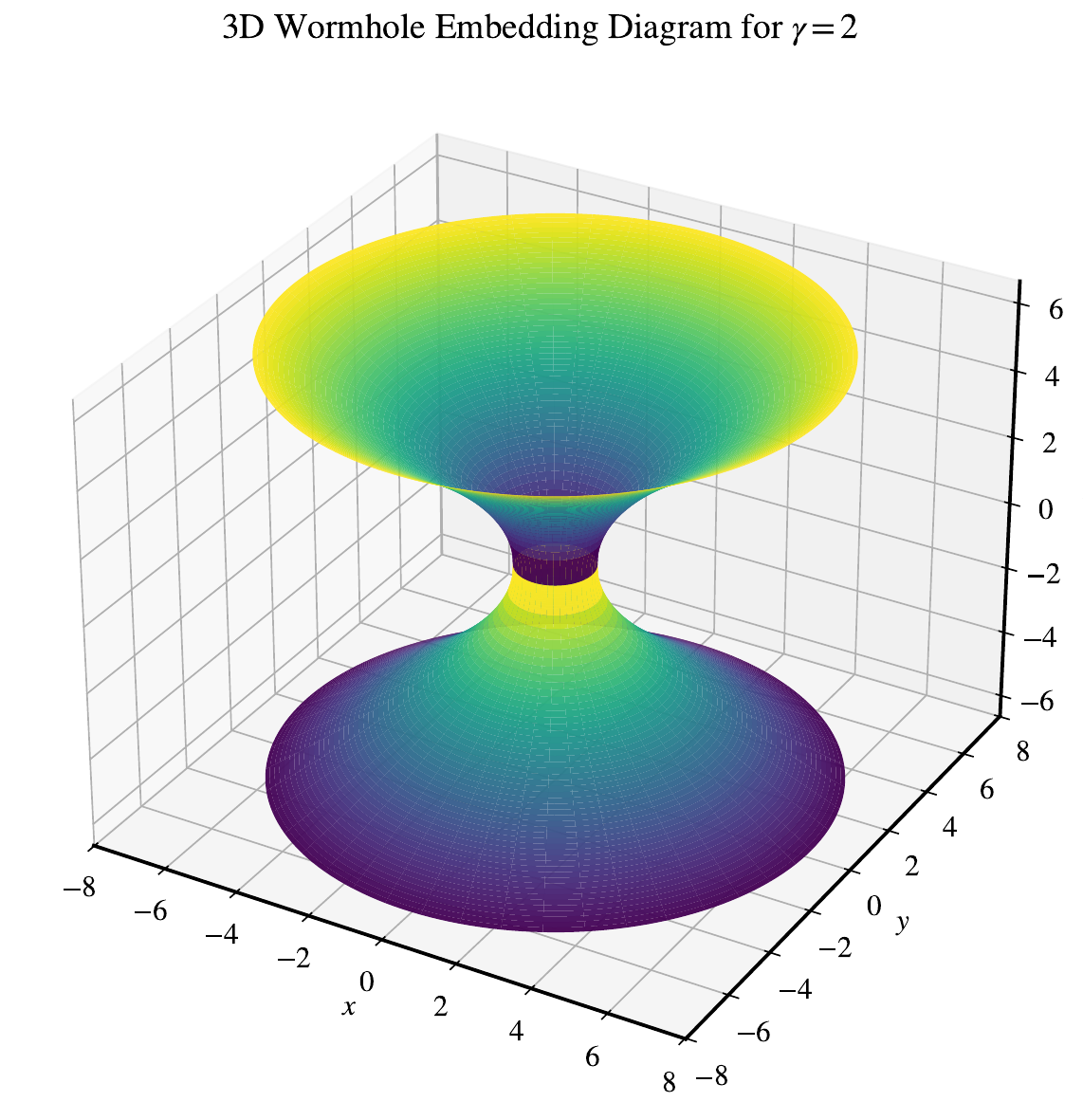}}}
\caption {Embeddings of the fractional HDE WHs for varying fractional parameters $\gamma$. The top and bottom embeddings show a smooth bridge between two asymptotically flat universes as a WH structure, with the change in the fractional parameter $\gamma$ affecting the structure without influencing the traversability of the WH.}\label{3f}
\end{figure}
The integration of Eq.~(1) is implemented using a reliable Python library and is shown in Fig. \textbf{\ref{3f}}. The upper branch of the embedding surface is shown in Fig. \textbf{\ref{3f}}(a), while Fig. \textbf{\ref{3f}}(b) presents the lower branch of the embedding surface. They join at the throat $r=r_{0}$, forming a continuous bridge between two asymptotically flat regions.

The embedding functions are finite and continuously differentiable throughout the integration domain, indicating that no geometrical singularities appear away from the throat. As the radial coordinate increases, both branches gradually flatten, reflecting the asymptotically Euclidean nature of the spacetime.

The complete three-dimensional embedding surface is shown in Fig. \textbf{\ref{3f}}(c). The familiar tunnel-like geometry characteristic of MT WHs is clearly observed. The throat represents the narrowest cross-section connecting two identical asymptotically flat universes. Variations of the fractional parameter modify the width and opening rate of the tunnel only slightly, while the global topology of the WH remains unchanged.

Therefore, the embedding analysis confirms that the obtained fractional HDE solution represents a smooth, traversable WH geometry satisfying all geometrical requirements, including the throat condition, flare-out criterion, and asymptotic flatness.

\section{Thermodynamic Properties of the Wormhole}
The thermodynamics of the HDE WH model can be studied using parameters such as the Hawking temperature, WH temperature, effective pressure, internal energy, work density, energy flux density, and specific heat capacity. All these physical parameters play a crucial role in describing the thermodynamics of the WH model. Throughout the analysis, we fix $d=0.05$, $r_{0}=1$, and study the influence of the fractional parameter $\gamma$.
\subsection{Hawking Temperature}
The Hawking temperature is a basic quantity in the analysis of traversable WHs from a thermodynamical point of view, because it gives information about the thermal characteristics of the system related to the WH horizon or throat. This quantity gives information about the heat exchange that exists between the WH configuration and the ambient spacetime. Hawking temperature in relation to the WH solution is defined as \cite{keathley2022detectability}
\begin{equation}
\mathcal{T}_\mathcal{HK}=\frac{\sqrt{1-\Psi(r)/r}}{2\pi}\phi'(r),
\end{equation}
For the chosen expression for the redshift function and the established form of the shape function, the Hawking temperature takes on the following value
\begin{equation}
\mathcal{T}_\mathcal{HK}=
\frac{0.0159155}{r^{2}}
\sqrt{
1-
\frac{
12\pi d^{2}\gamma r^{2/\gamma}
+r_{0}
-
12\pi d^{2}\gamma r_{0}^{2/\gamma}
}{r}
}.
\end{equation}
\begin{figure}[H]
\centering
\subfloat[]{{\includegraphics[height=2 in, width=3 in]{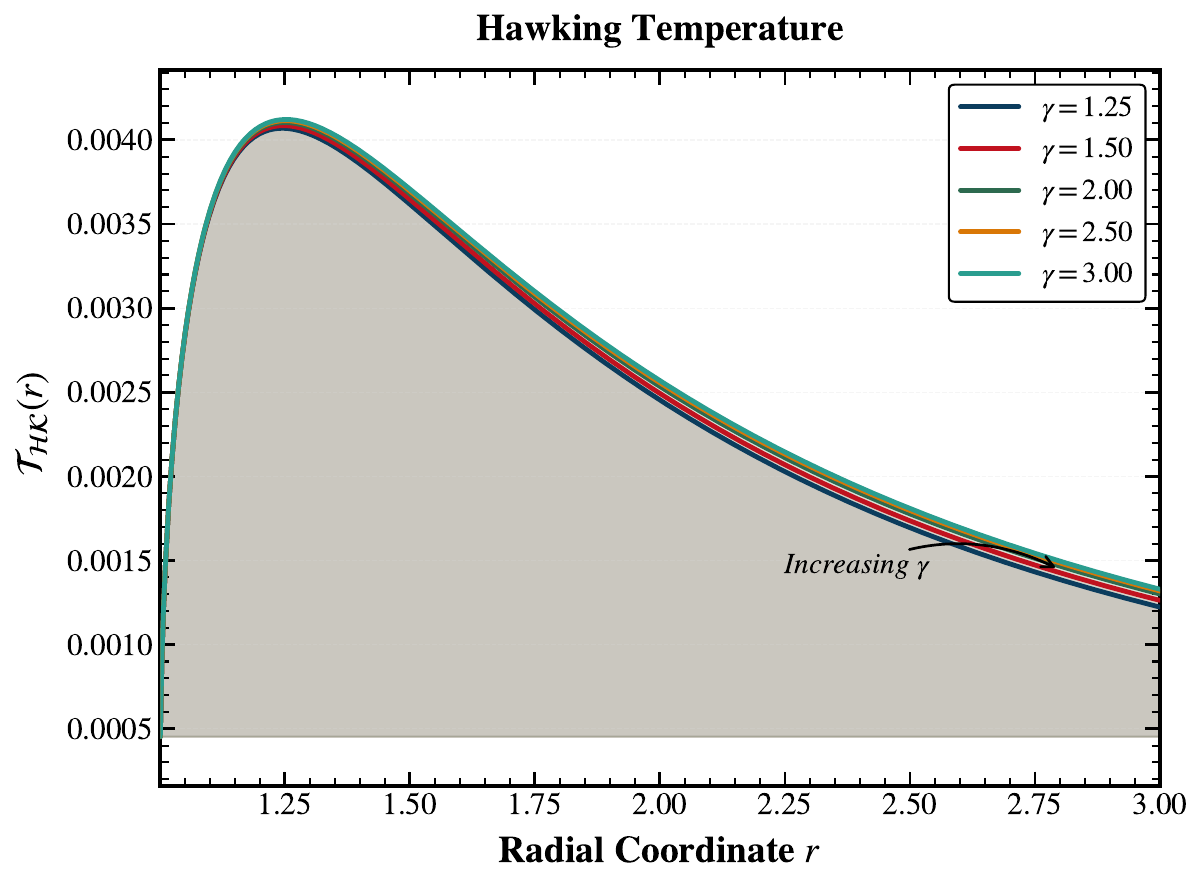}}}
\qquad
\subfloat[]{{\includegraphics[height=2 in, width=3 in]{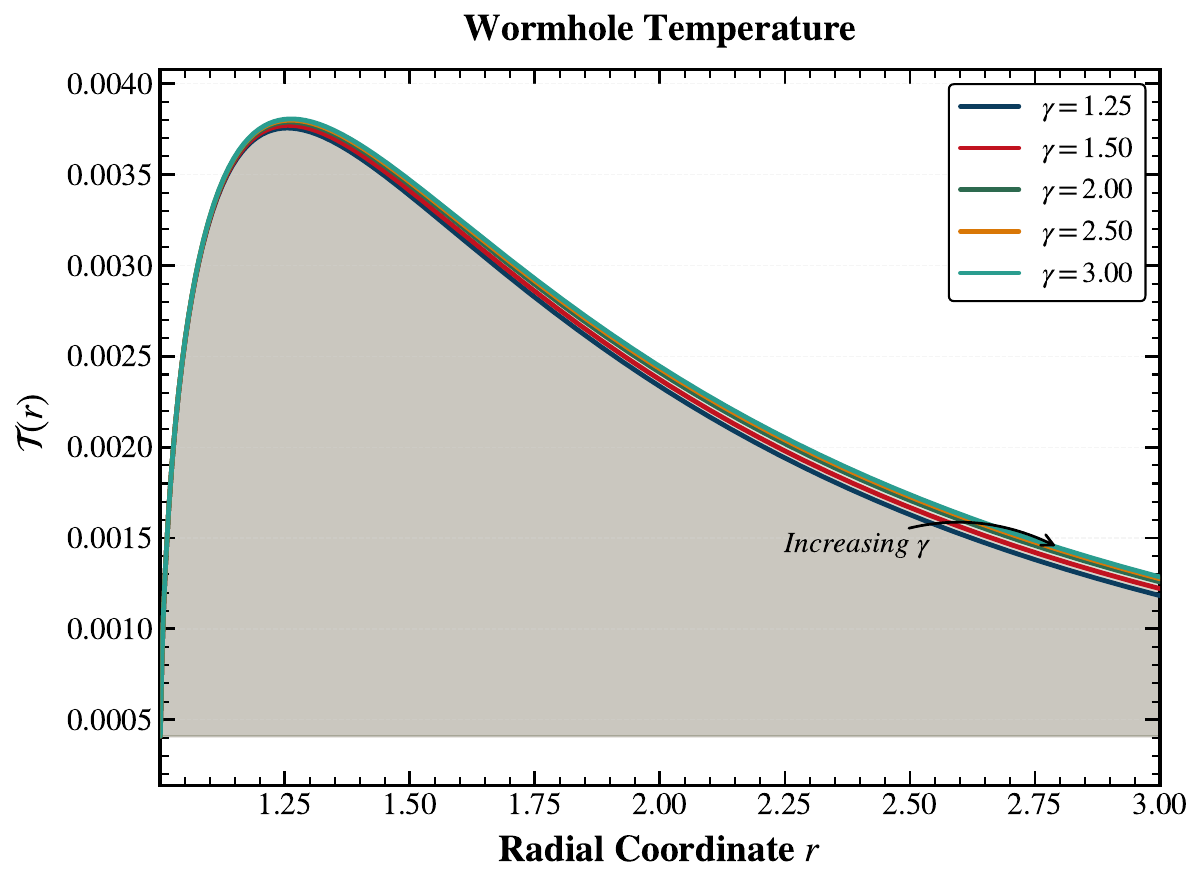}}}
\qquad
\subfloat[]{{\includegraphics[height=2 in, width=3 in]{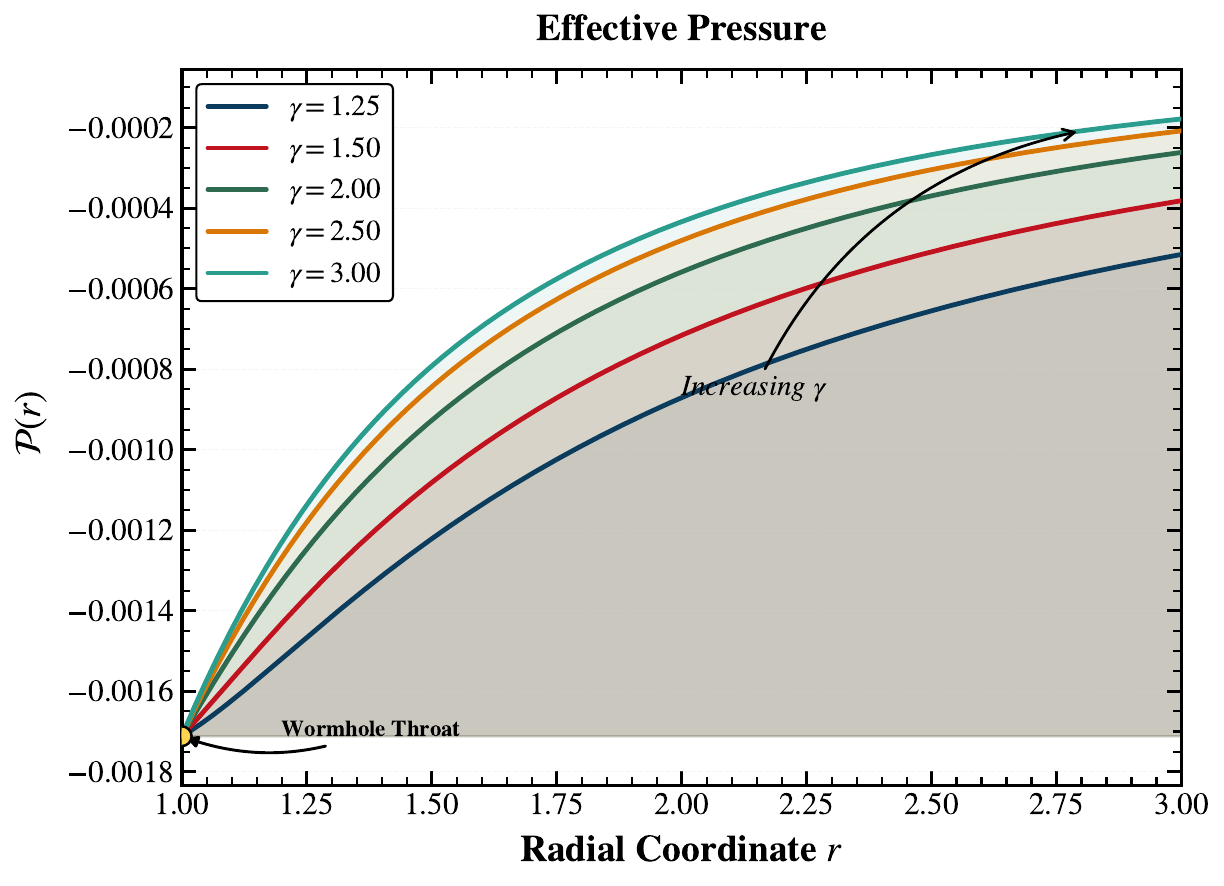}}}
\caption {The variation of the Hawking temperature, wormhole temperature, and effective pressure with the fractional holographic dark energy wormhole for $c=0.05$ and $r_0=1$ with various values of the fractional parameter $\gamma$. The fractional parameter affects the thermodynamic functions considerably while still preserving well-behaved results.}\label{4f}
\end{figure}
Hawking temperature for different fractional parameter values is shown in Fig. \textbf{\ref{4f}}(a). The temperature is positive and decreases with an increase in the radial coordinate. It attains its maximal value close to the WH throat and tends to zero when we approach infinity. Thus, spacetime becomes colder at infinity \cite{gonzalez2012thermal,hong2006can}. Variation in the value of $\gamma$ affects only the value of the temperature but not its behavior, which proves the thermodynamic regularity of the constructed WHs.
\subsection{Wormhole Temperature}
Wormhole temperature takes into account the gravitational redshift according to the Tolman formula and is the local temperature of the WH space-time. This temperature gives a better representation of the WH's thermal state and also gives important information on the thermal behavior of the WH under the action of the matter that sustains it. It is given by
\begin{equation}
\mathcal{T}(r)=e^{\phi(r)}\mathcal{T}_\mathcal{HK},
\end{equation}
which yields
\begin{equation}
\mathcal{T}(r)=
\frac{0.0159155e^{-0.1/r}}{r^{2}}
\sqrt{
1-
\frac{
12\pi c^{2}\gamma r^{2/\gamma}
+r_{0}
-
12\pi c^{2}\gamma r_{0}^{2/\gamma}
}{r}
}.
\end{equation}
This behavior of the WH temperature can be seen in Fig. \textbf{\ref{4f}}(b). Similarly to the Hawking temperature, the temperature of the WH is a smooth decreasing function from the throat to the asymptotic region due to the exponential redshift effect. The temperature is always finite and positive for any possible value of the parameter $\gamma$, demonstrating that the fractional HDE WH possesses a physically acceptable thermal profile.

The effective pressure is given by \cite{ma2015stability}
\begin{equation}
\mathcal{P}(r)=\frac{p_{rd}+2p_{tn}}{3}.
\end{equation}
From Fig. \textbf{\ref{4f}}(c), the radial profile of the effective pressure can also be seen. In general, the effective pressure is well-behaved in the whole spacetime without any kind of discontinuity or singularity. The effective pressure decreases monotonically with an increase in the value of the radial coordinate and becomes almost zero at large distances from the throat region. The sensitivity to the fractional parameter is quite weak, implying that the fractional holographic corrections have a little effect on the value of the pressure while preserving the overall physical behavior of the WH matter distribution.

\subsection{Internal Energy, Work Density and Energy Flux}
The internal energy refers to the energy associated with the WH due to its mass content and the nature of the space-time, thereby making it one of the most important parameters in determining the physical viability of the WH from a thermodynamic perspective. Positive and finite internal energy, together with a smooth increase in the same, signifies that the WH is still physically viable within the entire spacetime region under consideration. The internal energy of the WH comes from the first law of WH thermodynamics \cite{saiedi2012thermodynamics}.
\begin{equation}
dE=\mathcal{T}_\mathcal{HK}dS+WdV,
\end{equation}
where
\begin{equation}
S=4\pi r^2,
\end{equation}
and
\begin{equation}
W=\frac{\mu-p_{rd}}{2}
\end{equation}
is the work density.

After substituting the entropy $S$ and volume element $V$, one obtains
\begin{equation}
\frac{dE}{dr}
=
16\pi r\mathcal{T}_\mathcal{HK}
+
4\pi r^2W,
\end{equation}
whose integration determines the total internal energy of the WH.
\begin{figure}[H]
\centering
\subfloat[]{{\includegraphics[height=2 in, width=3 in]{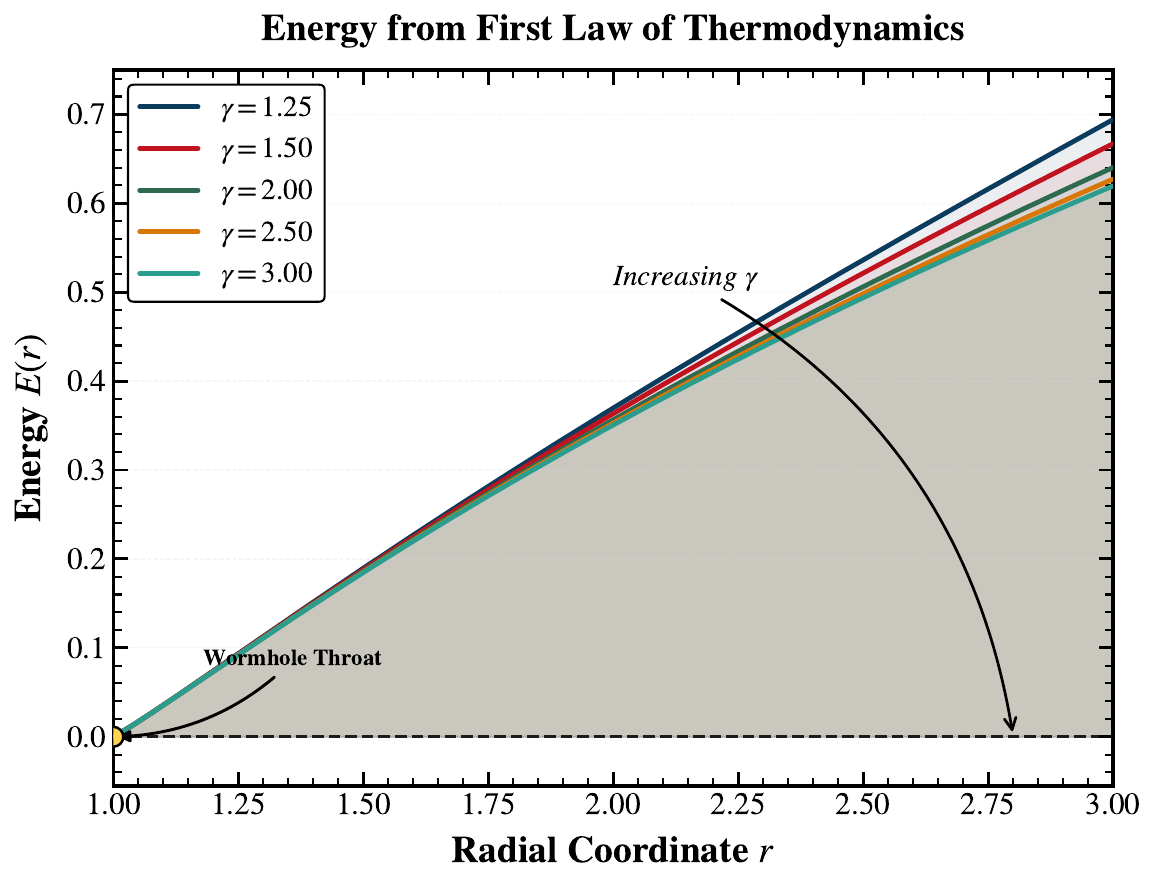}}}
\qquad
\subfloat[]{{\includegraphics[height=2 in, width=3 in]{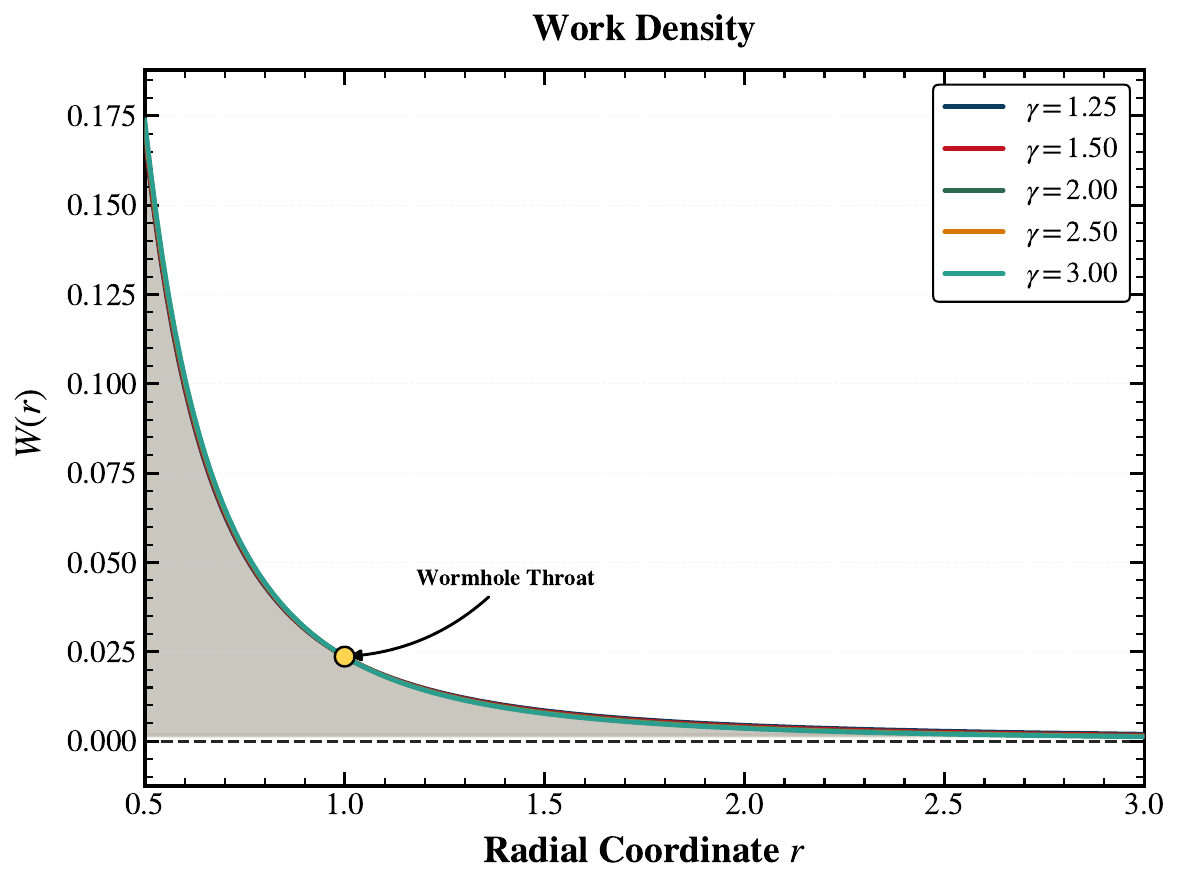}}}
\qquad
\subfloat[]{{\includegraphics[height=2 in, width=3 in]{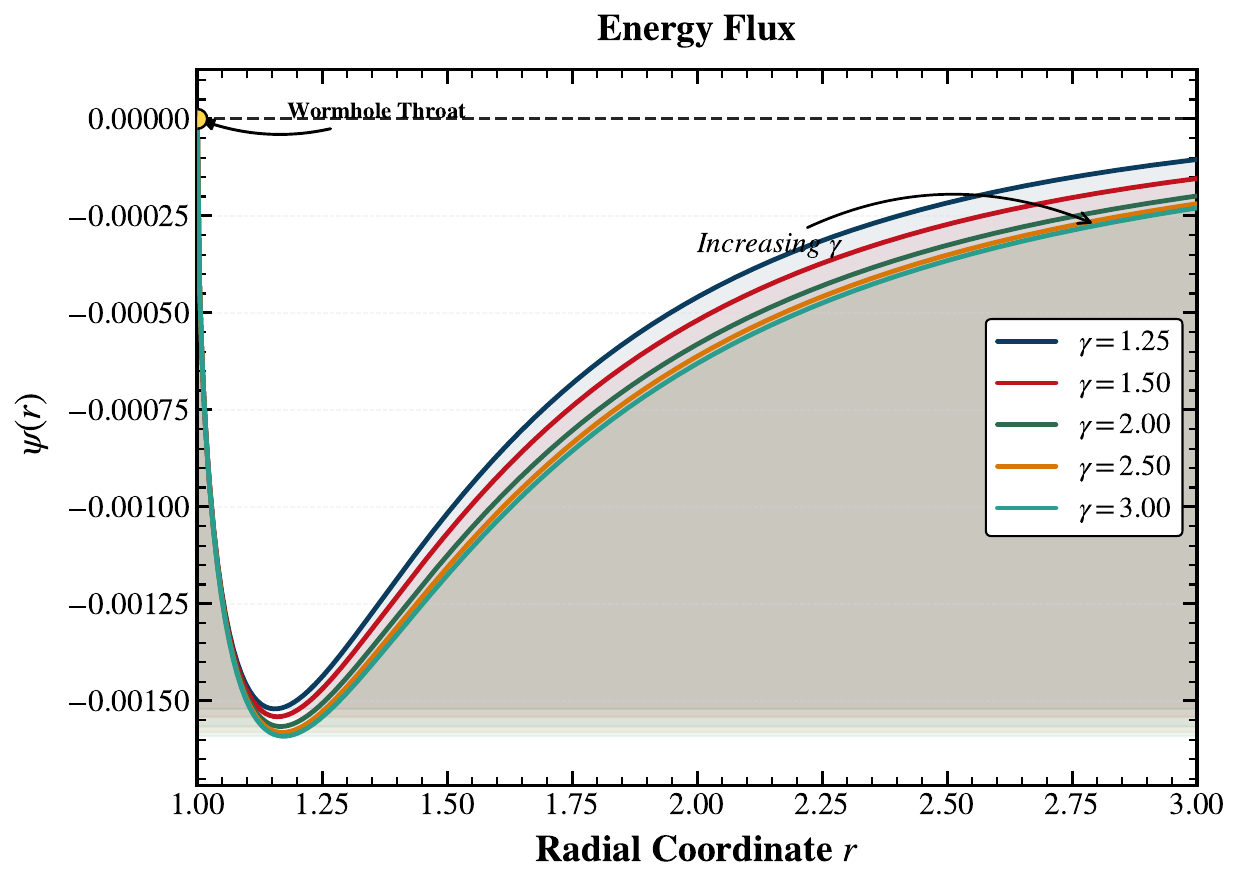}}}
\qquad
\subfloat[]{{\includegraphics[height=2 in, width=3 in]{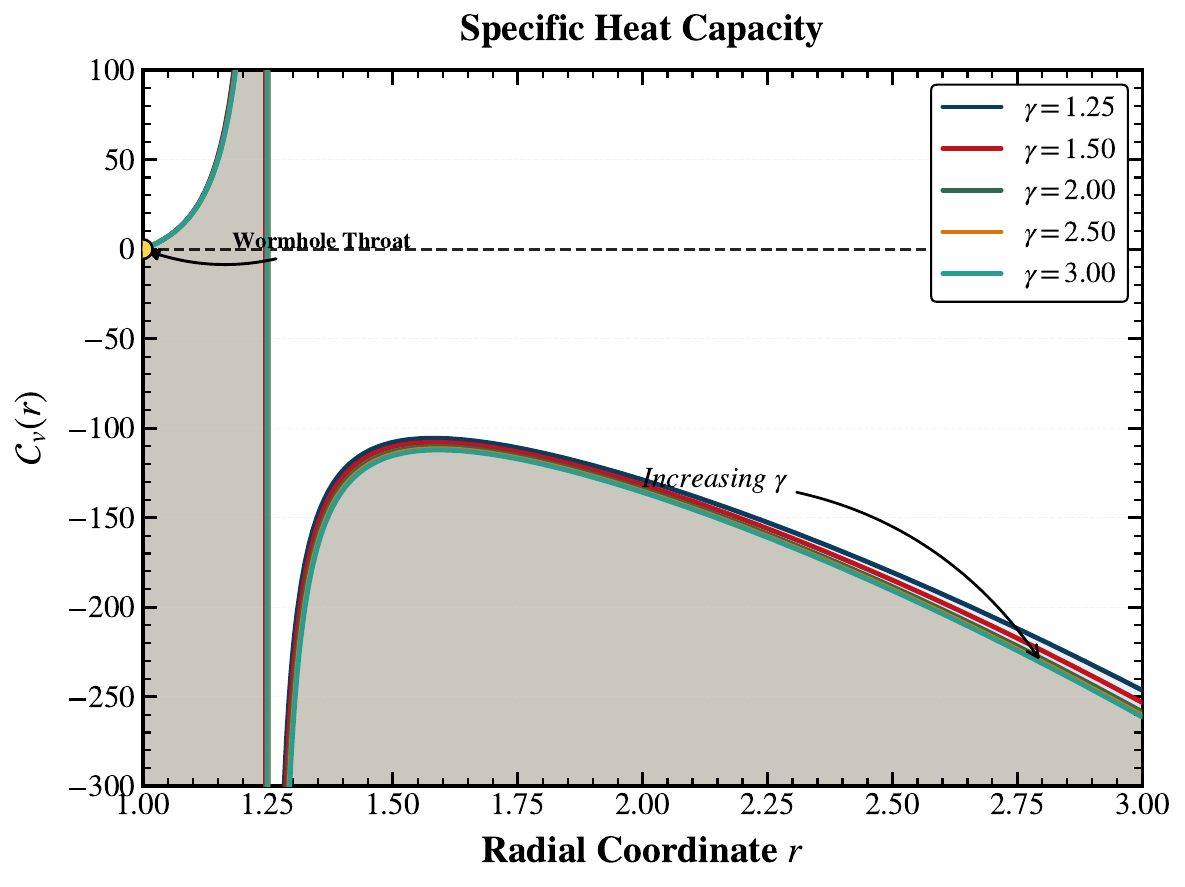}}}
\caption {Profile of internal energy evaluated by the application of the first law of thermodynamics, work density, energy flux, and specific heat capacity of fractional HDE WHs for $d = 0.05$, $r_0 = 1$, and different values of the fractional parameter $\gamma$.}\label{5f}
\end{figure}
The result in Figure~\ref{5f}(a) implies that the internal energy always grows from the throat region outwards. The monotonic growth pattern signifies the fact that the presence of fractonal HDE matter adds up to the thermodynamic energy of the WH solution. The influence of the fractional parameter is evident for large radial coordinates,  where larger values of $\gamma$ produce slightly different growth rates without changing the qualitative behavior.

The work density is defined as
\begin{equation}
W=\frac{\mu-p_{rd}}{2},
\end{equation}
and gives an estimation of work done by the effective matter source on the spacetime geometry. The work density, as seen from Fig.~\ref{5f}(b), is always finite and positive within the whole spacetime. The work density decreases as the radial coordinate increases, showing that the effective matter does most of the work close to the throat region, while the contribution gradually diminishes in the asymptotic region.

The corresponding vector of energy supply (energy flux) is written as \cite{martin2011lorentzian,martin2009thermal}
\begin{equation}
\psi=
\frac{e^{\phi(r)}}{4}
(\mu+p_{rd})
\sqrt{1-\frac{\Psi(r)}{r}}.
\end{equation}
As can be seen from the corresponding plots in Fig. \textbf{\ref{5f}}(c), the energy flux attains its maximal value at the throat and gradually drops with increasing $r$. This implies that the energy transport occurs mostly near the throat of the WH, while the energy transport in the asymptotically flat spacetime is insignificant. The fractional parameter mainly affects the magnitude of the energy flux while keeping its radial variation unchanged.

\subsection{Specific Heat Capacity}
Thermodynamic stability of the WH solution is considered in terms of the specific heat capacity \cite{callen1985thermostatistics,ma2015stability,ditta2025exploring},
\begin{equation}
\mathcal{C}_v =
\mathcal{T}_\mathcal{HK} \frac{\partial S / \partial r}
{\partial \mathcal{T}_\mathcal{HK} / \partial r}.
\end{equation}
In Fig. \textbf{\ref{5f}}(d), the plots of the specific heat capacity are presented. They are well-defined in all physically possible regions. The positive values of $\mathcal{C}_v$ correspond to the thermodynamically stable configurations, while negative values mean the unstable states. For the chosen parameters, $\mathcal{C}_v$ is mostly positive, which means that the fractional HDE WH is thermodynamically stable in most of the spacetime, especially near the WH throat. The increase of the fractional parameter does not change the behavior of $\mathcal{C}_v$, but only changes the magnitude of the corresponding quantities.

In summary, the thermodynamic properties demonstrate the regularity of thermal behavior of the fractional HDE WH. Hawking and WH temperatures are positive, the effective pressure changes regularly, the internal energy grows monotonically, the work density and energy flux are finite, and the positive value of the specific heat capacity near the WH throat indicates the thermodynamic stability of the WH.

\section{Physical Analysis of the Wormhole Solutions}
After establishing the geometrical viability of the obtained fractional HDE WH, we now investigate its physical properties. In particular, we examine the energy conditions, anisotropic behavior, EoS, active gravitational mass, compactness, exoticity parameter, volume integral quantifier, and equilibrium of the matter distribution. These analyses provide important information regarding the physical acceptability and stability of the obtained WH configuration.
\subsection{Energy Conditions and Anisotropy}

Energy conditions form a key criterion for evaluating the physical consistency of the matter content that forms the basis for a traversable WH solution. Among the energy conditions, the NEC plays a crucial role since its non-satisfaction is the basic necessity for an open WH mouth in GR \cite{morris1988wormholess,visser1995lorentzian,hochberg1998null}. Strong energy condition (SEC) identifies the overall gravitational nature of the matter content. The anisotropy parameter represents the difference between radial and tangential pressure and is responsible for providing extra tension to sustain the WH structure.

Radial and tangential NEC and SEC can be expressed mathematically as
\begin{equation}
\mu+p_{rd}\ge0,~~ \mu+p_{tn}\ge0,
\end{equation}
and
\begin{equation}
\mu+p_{rd}+2p_{tn}\ge0,
\end{equation}
respectively, whereas the anisotropy parameter is given by
\begin{equation}
\Delta=p_{tn}-p_{rd}.
\end{equation}

\begin{figure}[H]
\centering
\subfloat[]{{\includegraphics[height=2 in, width=3 in]{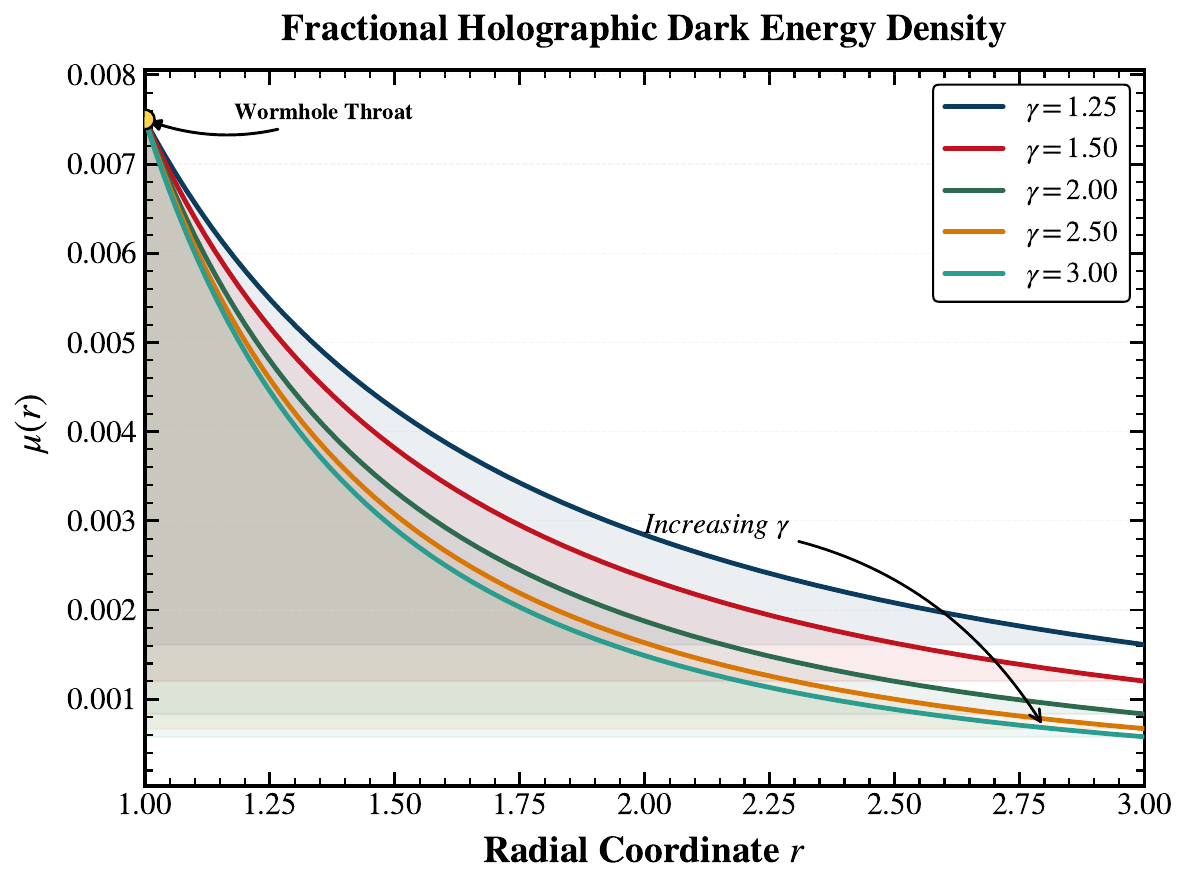}}}
\qquad
\subfloat[]{{\includegraphics[height=2 in, width=3 in]{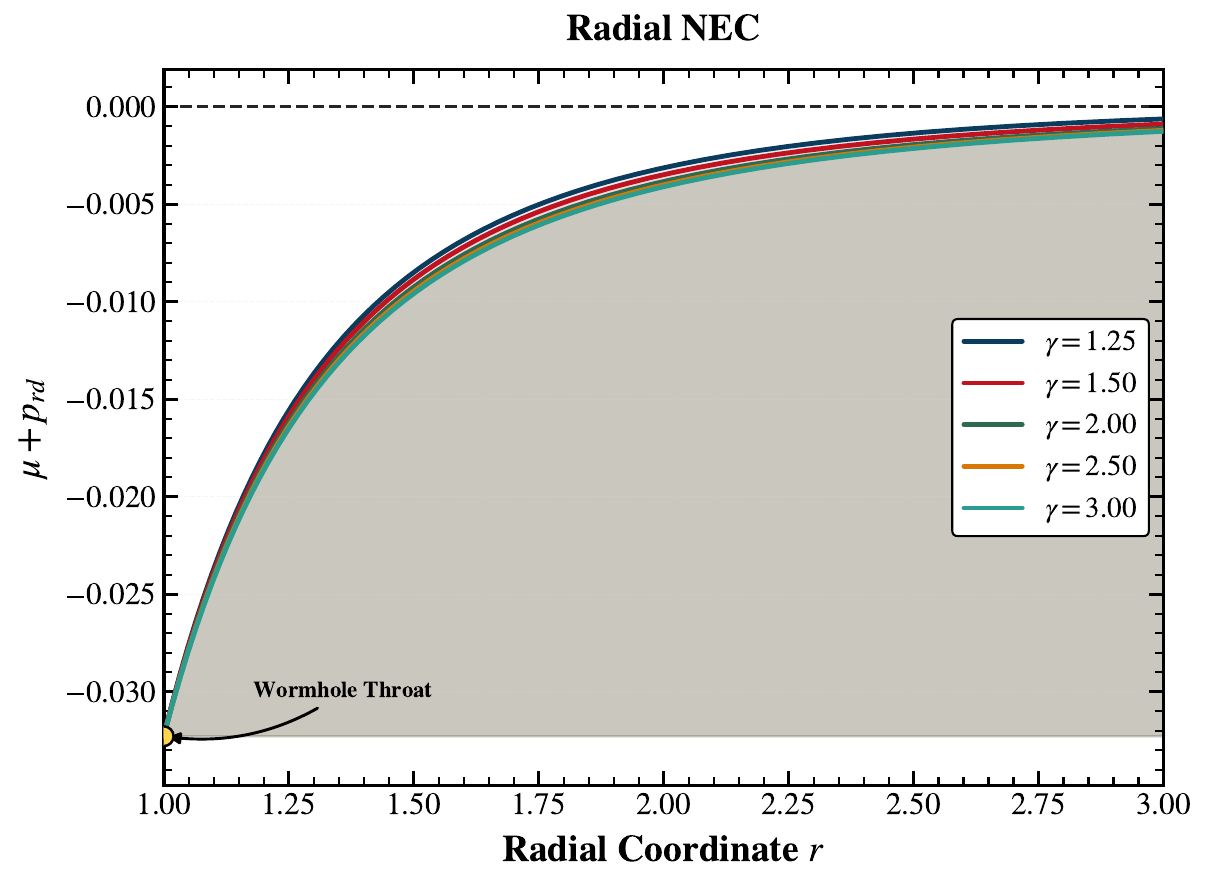}}}
\qquad
\subfloat[]{{\includegraphics[height=2 in, width=3 in]{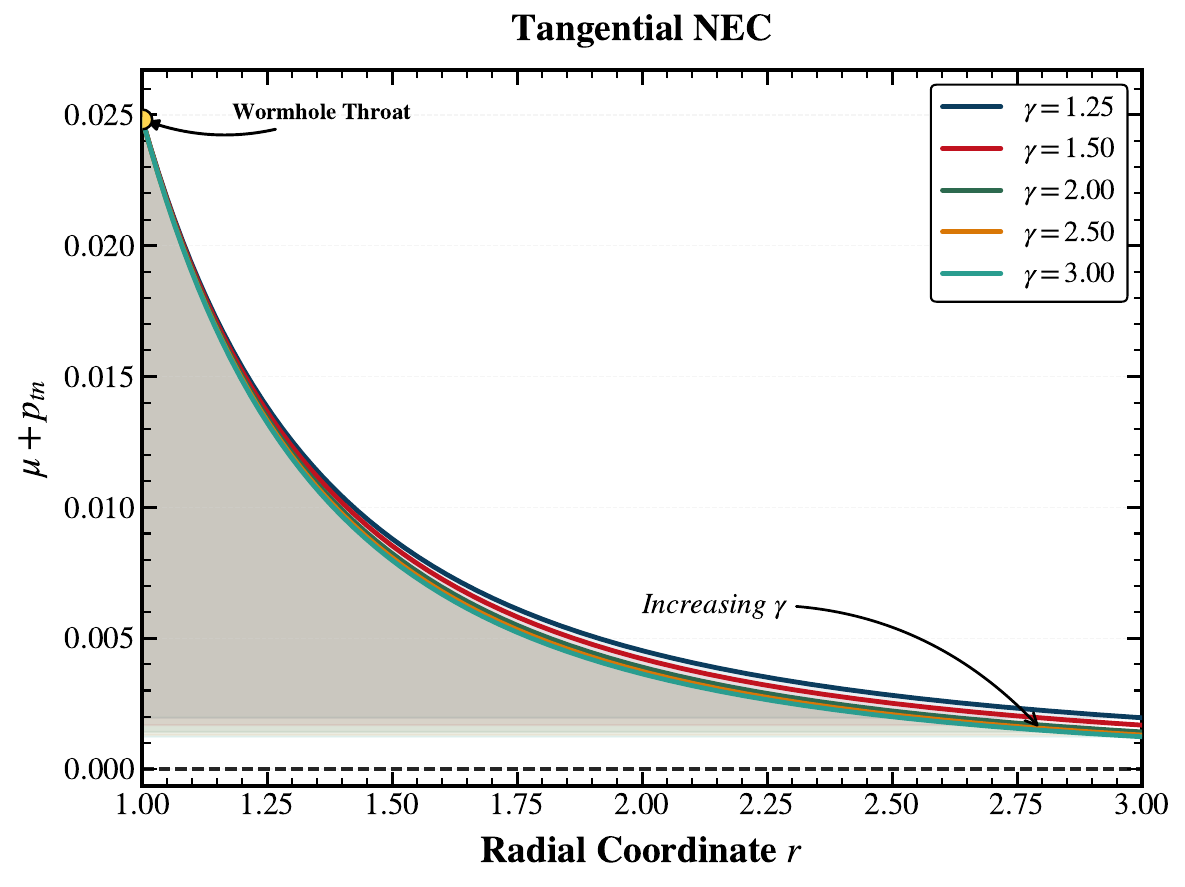}}}
\qquad
\subfloat[]{{\includegraphics[height=2 in, width=3 in]{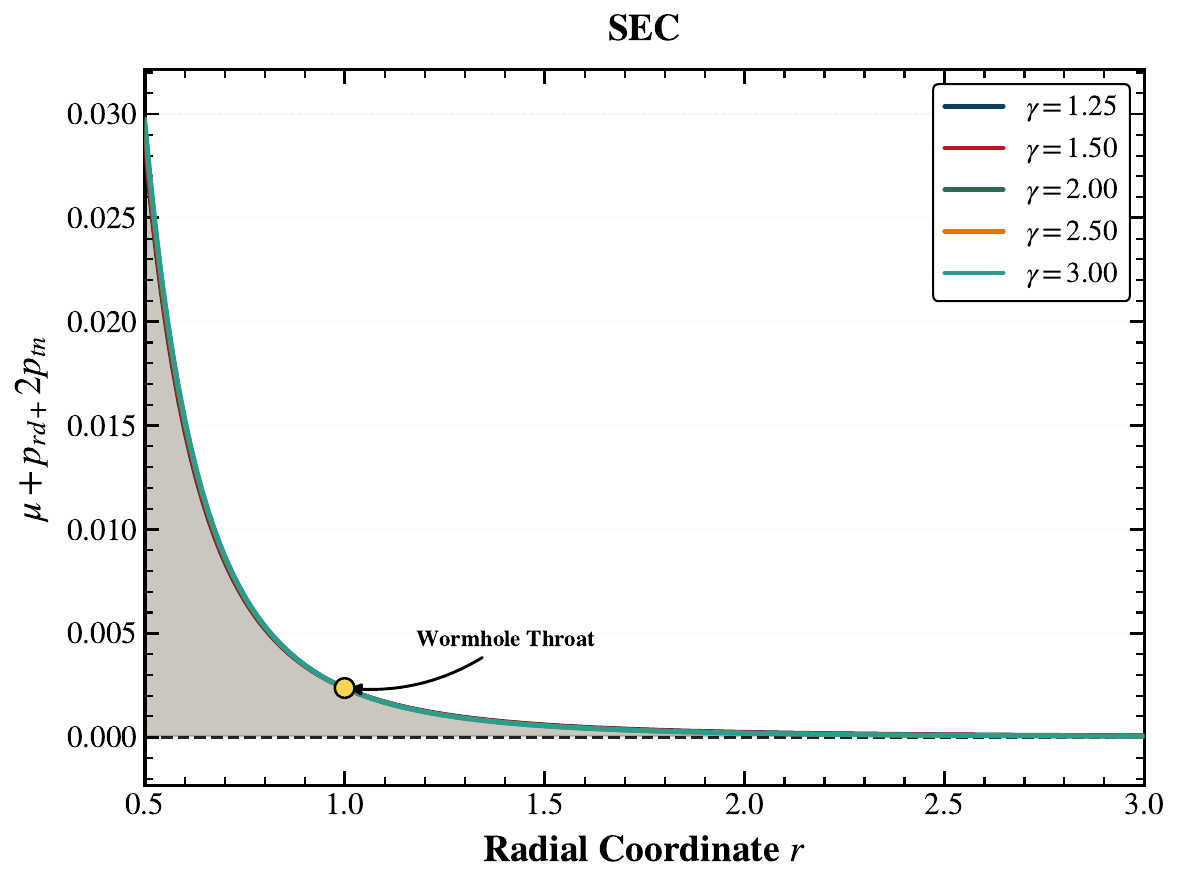}}}
\qquad
\subfloat[]{{\includegraphics[height=2 in, width=3 in]{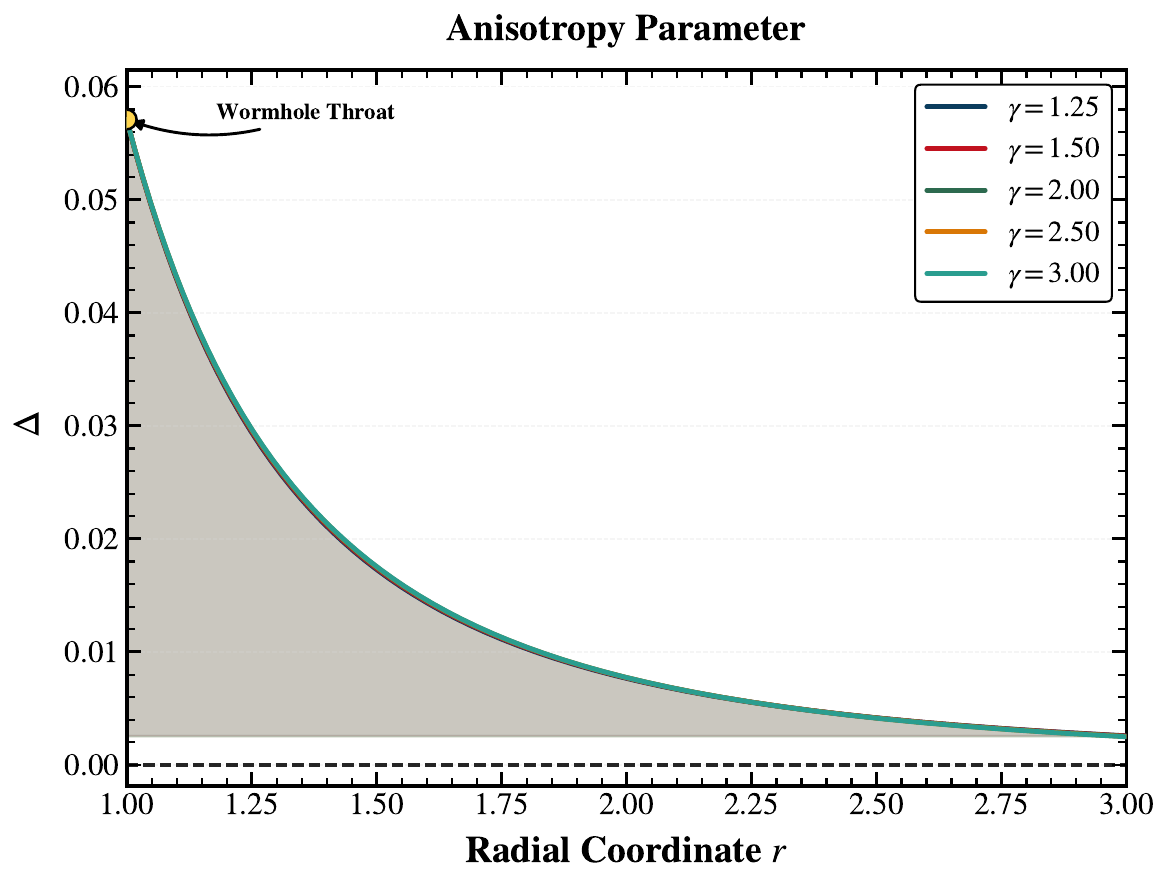}}}
\caption {Profiles of the energy density, energy conditions, and anisotropy for the fractional HDE WH corresponding to different values of the fractional parameter $\gamma$. The results show a positive energy density, violation of only the radial NEC near the throat, and physically acceptable behavior of the remaining quantities.}\label{6f}
\end{figure}

From the graphical interpretation provided by Fig. \textbf{\ref{6f}}, it becomes clear that the radial NEC is violated within the vicinity of the throat of the WH for all values of the fractional parameter $\gamma$. On the other hand, the tangential NEC remains valid within the complete domain. The violation in this case indicates the existence of the exotic matter needed to make a traversable WH. The intensity of violation decreases as the radial distance increases, indicating that the exotic behavior is primarily localized near the throat.

On the other hand, the SEC is quite happy with the given domain, which shows that even though the energy density distribution fails to satisfy the radial NEC, it is satisfied by the combined condition of $\mu+p_{rd}+2p_{tn}\ge0$. This fact reveals that the fractional HDE only gives the least possible exotic nature needed to maintain the WH geometry without causing any large violation of the usual energy conditions.

The value of the anisotropy parameter remains positive everywhere in the space-time region, indicating that the tangential pressure is greater than the radial pressure. As a result, the anisotropic pressure pushes the WH neck outward and plays an important role in preventing the WH neck from collapsing \cite{herrera1997local,bowers1974anisotropic}. Also, when the value of the fractional parameter $\gamma$ increases, the degree of violation of the radial NEC decreases and the anisotropic stress becomes less, suggesting that larger values of $\gamma$ lead to comparatively less exotic and more physically acceptable WH configurations while preserving traversability. This behavior is also evident from Table \textbf{\ref{tab1}}.
\begin{table}[H]
\caption{\label{tab1} Summary of the physical behavior of the fractional HDE WH solution.}
\begin{ruledtabular}
\begin{tabular}{lcccl}
Quantity &
Near throat &
Away from throat &
Status &
Physical interpretation\\
\hline

Energy density $\mu$
&
Positive
&
Positive
&
Physical
&
Positive matter distribution.
\\

Radial NEC $(\mu+p_{rd})$
&
Violated
&
Weak violation
&
Violated
&
Supports traversable throat.
\\

Tangential NEC $(\mu+p_{tn})$
&
Satisfied
&
Satisfied
&
Satisfied
&
Regular tangential stresses.
\\

SEC $(\mu+p_{rd}+2p_{tn})$
&
Satisfied
&
Satisfied
&
Satisfied
&
Minimal exotic matter.
\\

Anisotropy $\Delta$
&
Positive
&
Positive
&
Stable
&
Outward stabilizing force.
\\

Radial EoS $\mathcal{W}_{rd}$
&
$\mathcal{W}_{rd}<-1$
&
Approaches $-1$
&
Phantom
&
Exotic matter source.
\\

Tangential EoS $\mathcal{W}_{tn}$
&
Finite
&
Finite
&
Regular
&
Smooth pressure profile.
\\

\end{tabular}
\end{ruledtabular}
\end{table}

\subsection{Equation of State}

Parameters of the EoS describe the properties of the matter that supports the WH, allowing one to classify the source either as normal matter, quintessence, or phantom energy. Being the key feature of any traversable WH, the EoS is one of the most crucial diagnostics for classification of the source.

Radial and tangential EoS parameters are written as
\begin{equation}
\mathcal{W}_{rd}=\frac{p_{rd}}{\mu},
\end{equation}
and
\begin{equation}
\mathcal{W}_{tn}=\frac{p_{tn}}{\mu}.
\end{equation}
\begin{figure}[H]
\centering
\subfloat[]{{\includegraphics[height=2 in, width=3 in]{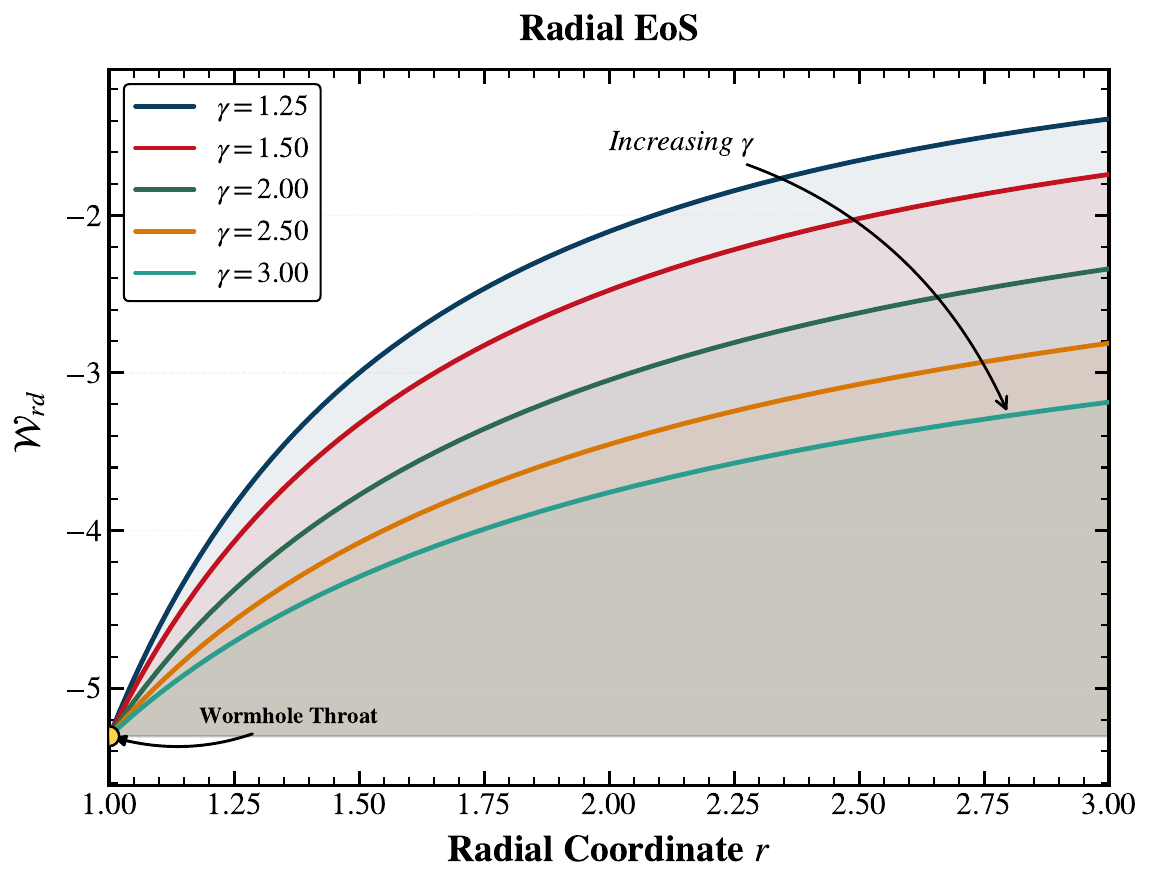}}}
\qquad
\subfloat[]{{\includegraphics[height=2 in, width=3 in]{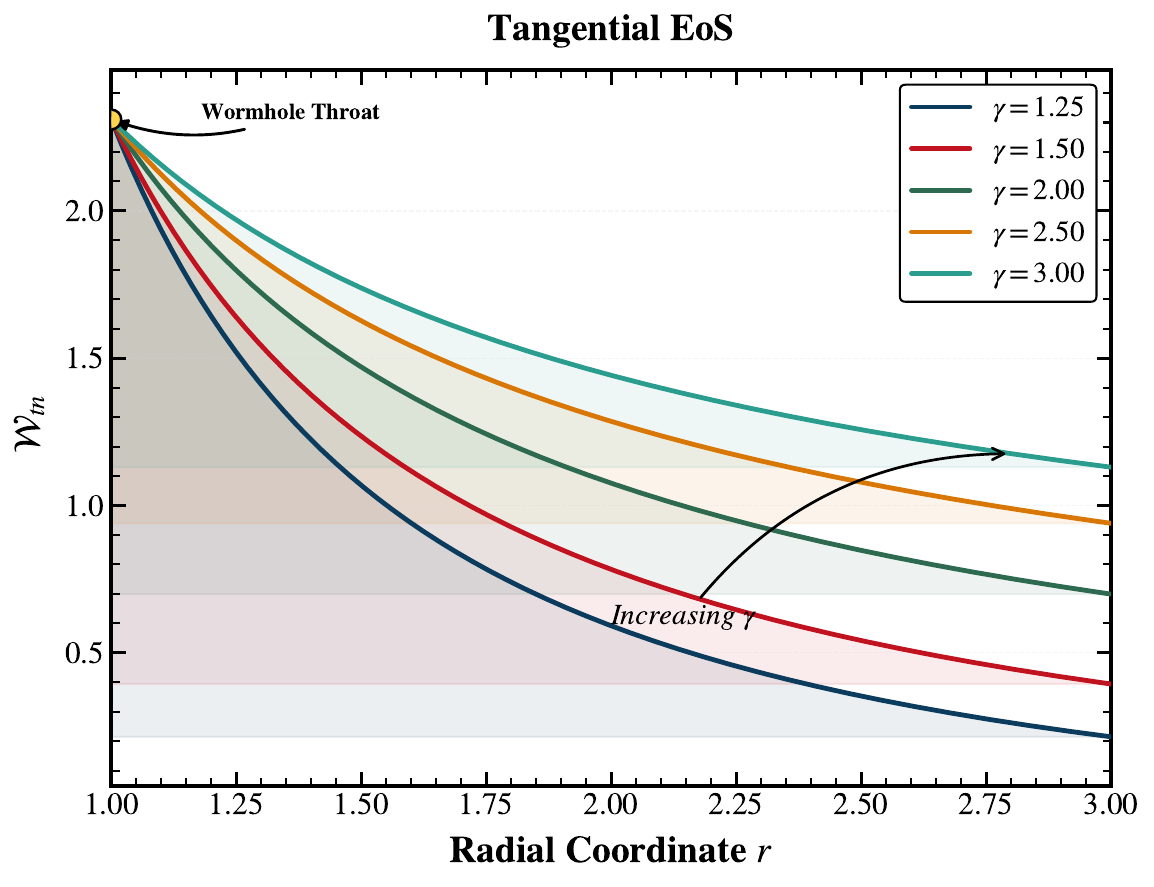}}}
\caption {Graphical visualization of the EoS parameters for the paramtric values $d=0.05$, $r_0=1$, and different fractional parameter $\gamma$.}\label{7f}
\end{figure}
Graphical behavior of the EoS parameters is depicted in Fig. \textbf{\ref{7f}}. The radial EoS parameter is always less than $-1$ near the throat, and hence the fractional HDE acts as a phantom fluid which can support the traversable WH. In turn, the tangential EoS parameter varies finitely without encountering any singularity.

The fractional parameter $\gamma$ governs the qualitative behavior of the EoS. With increasing $\gamma$, the values of both $\mathcal{W}_{rd}$ and $\mathcal{W}_{tn}$ tend to less negative values but still possess the phantom property, proving the fractional correction modifies only the strength of the exotic matter without changing its physical nature.

\subsection{Active Gravitational Mass and Compactness}

The active gravitational mass is the gravitating mass content up to a certain radius, which gives some insights into the gravitational nature of the WH. The compactness gives an idea about the gravitational compactness of the spacetime and is used to classify traversable WHs from black hole scenarios.

Active gravitational mass is defined by
\begin{equation}
M(r)=4\pi\int_{r_0}^{r}\mu(\tilde r)\tilde r^2\,d\tilde r,
\end{equation}
and the compactness is
\begin{equation}
U_{cp}(r)=\frac{M(r)}{r}.
\end{equation}
\begin{figure}[H]
\centering
\subfloat[]{{\includegraphics[height=2 in, width=3 in]{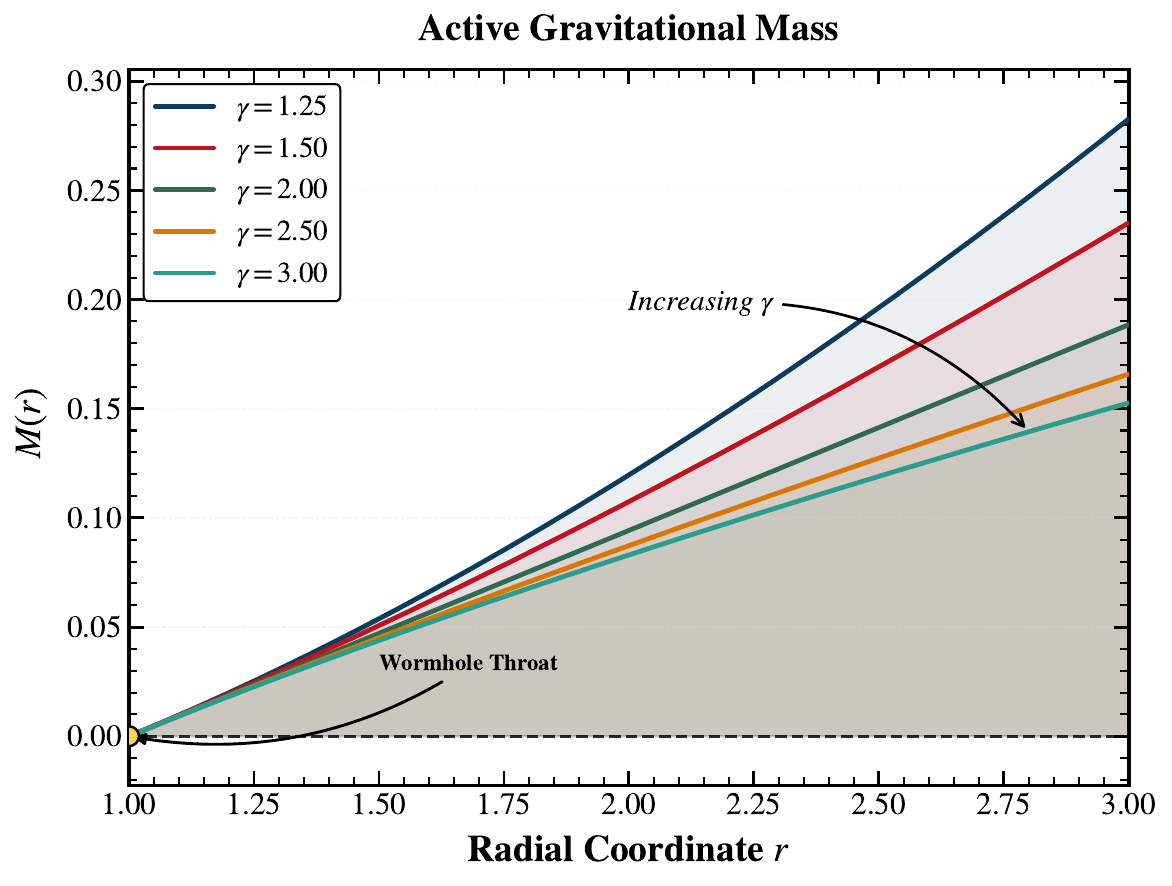}}}
\qquad
\subfloat[]{{\includegraphics[height=2 in, width=3 in]{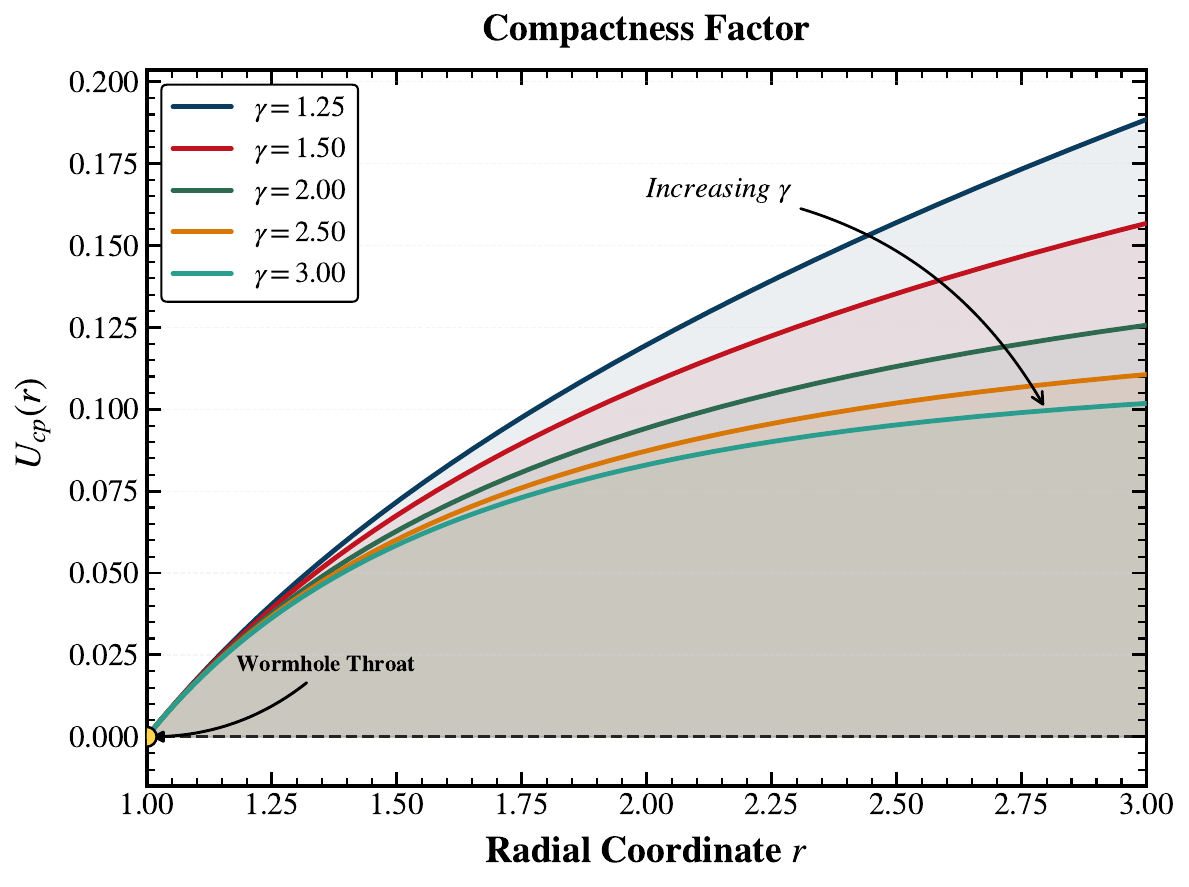}}}

\caption {Dynamics of the active gravitational mass and compactness of the fractional HDE WHs, corresponding to the parametric values $d=0.05$, $r_0=1$, and varying fractional parameter $\gamma$.}\label{8f}
\end{figure}
Graphical behavior of active gravitational mass and compactness is depicted in Fig. \textbf{\ref{8f}}. Solutions obtained have a positive, monotonically increasing active gravitational mass, which means that the WH has an increasing amount of mass far from the throat. Moreover, the compactness is finite for the physical range of coordinates, and it is always less than the black hole value, which tells us that there is no formation of an event horizon.

An increase in the fractional parameter $\gamma$ decreases the amount of gravitational mass and hence decreases the compactness.  Therefore, larger values of $\gamma$ lead to less compact WH geometries while preserving their overall physical viability.

\subsection{Exoticity Parameter and Volume Integral Quantifier}

The exoticity parameter gives the exact value of the amount of exotic matter involved in the WH spacetime geometry, while the volume integral quantifier represents an estimation of the total amount of exotic matter needed to maintain the WH. These measures are highly significant since it is expected for physically realizable traversable WHs to possess just a finite amount of exotic matter.

The expression for the exoticity parameter is \cite{morris1988wormholes,morris1988wormholess}
\begin{equation}
\chi=-\frac{\mu+p_{rd}}{|\mu|},
\end{equation}
whereas the formula for the volume integral quantifier is \cite{visser2003traversable}
\begin{equation}
I_V=
8\pi
\int_{r_0}^{\infty}
(\mu+p_{rd})r^2dr.
\end{equation}
\begin{figure}[H]
\centering
\subfloat[]{{\includegraphics[height=2 in, width=3 in]{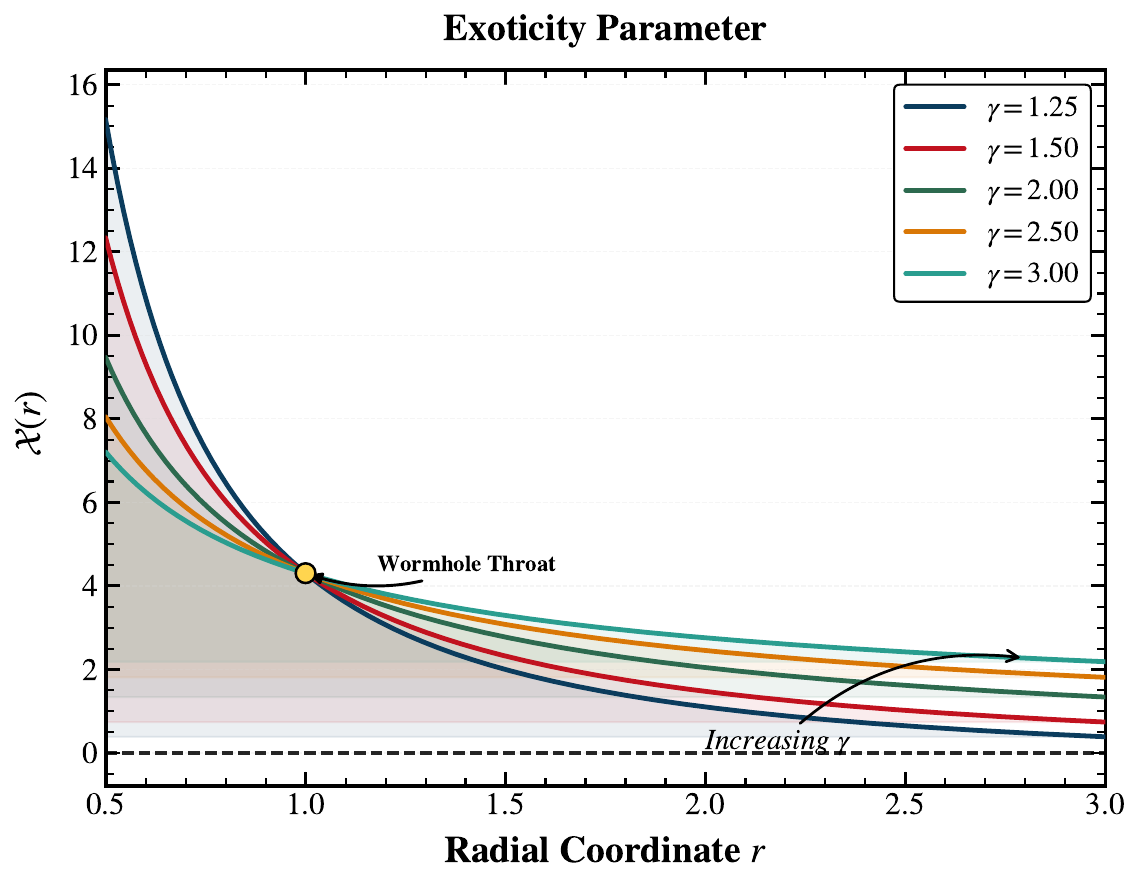}}}
\qquad
\subfloat[]{{\includegraphics[height=2 in, width=3 in]{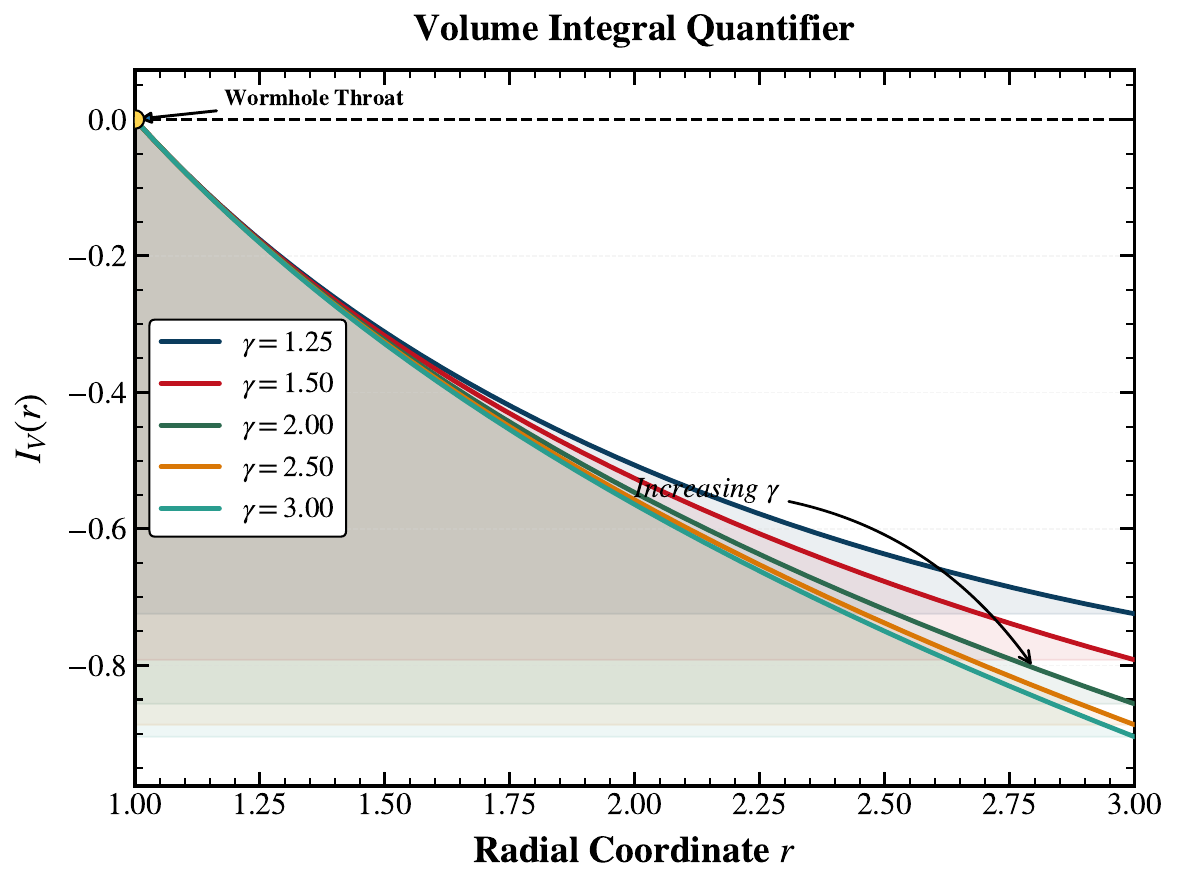}}}
\caption {Behavior of the exoticity parameter and volume integral quantifier for the fractional HDE WHs corresponding to the parametric values $d=0.05$, $r_0=1$, and different values of the fractional parameter $\gamma$. }\label{9f}
\end{figure}
The exoticity parameter and volume integral quantifier are demonstrated graphically in Fig. \textbf{\ref{9f}}. It can be observed that the exoticity parameter stays positive throughout the whole physically meaningful domain, which means that the matter source does indeed have the exotic properties necessary to keep the WH open. Besides, the exoticity parameter gets smaller with an increase in the radial coordinate, which means that the exotic nature of the matter source is confined in the throat neighborhood.

The volume integral quantifier takes a finite value for all chosen values of $\gamma$, meaning that it is indeed a finite amount of exotic matter that is required to keep the WH open. The greater the fractional parameter, the smaller the values of the exoticity parameter and the volume integral, suggesting that stronger fractional effects lessen the total exotic matter required to maintain the WH geometry.

\subsection{Equilibrium Analysis}

The equilibrium analysis checks the validity of the static nature of the WH matter configuration under the influence of gravitational, hydrostatic, and anisotropic forces. The existence of a traversable WH demands that the forces be balanced in the spacetime.

The generalized TOV (Tolman--Oppenheimer--Volkoff) equation (presented in equation \eqref{conservation}) can be written as \cite{tolman1939static}
\begin{equation}
\mathcal{F}_{gr}+\mathcal{F}_{hd}+\mathcal{F}_{an}=0,
\end{equation}
with
\begin{equation}
\mathcal{F}_{gr}=-\frac{\phi'}{2}(\mu+p_{rd}),\qquad
\mathcal{F}_{hd}=-\frac{dp_{rd}}{dr},\qquad
\mathcal{F}_{an}=\frac{2(p_{tn}-p_{rd})}{r}.
\end{equation}
\begin{figure}[H]
\centering
\subfloat[]{{\includegraphics[height=2 in, width=3 in]{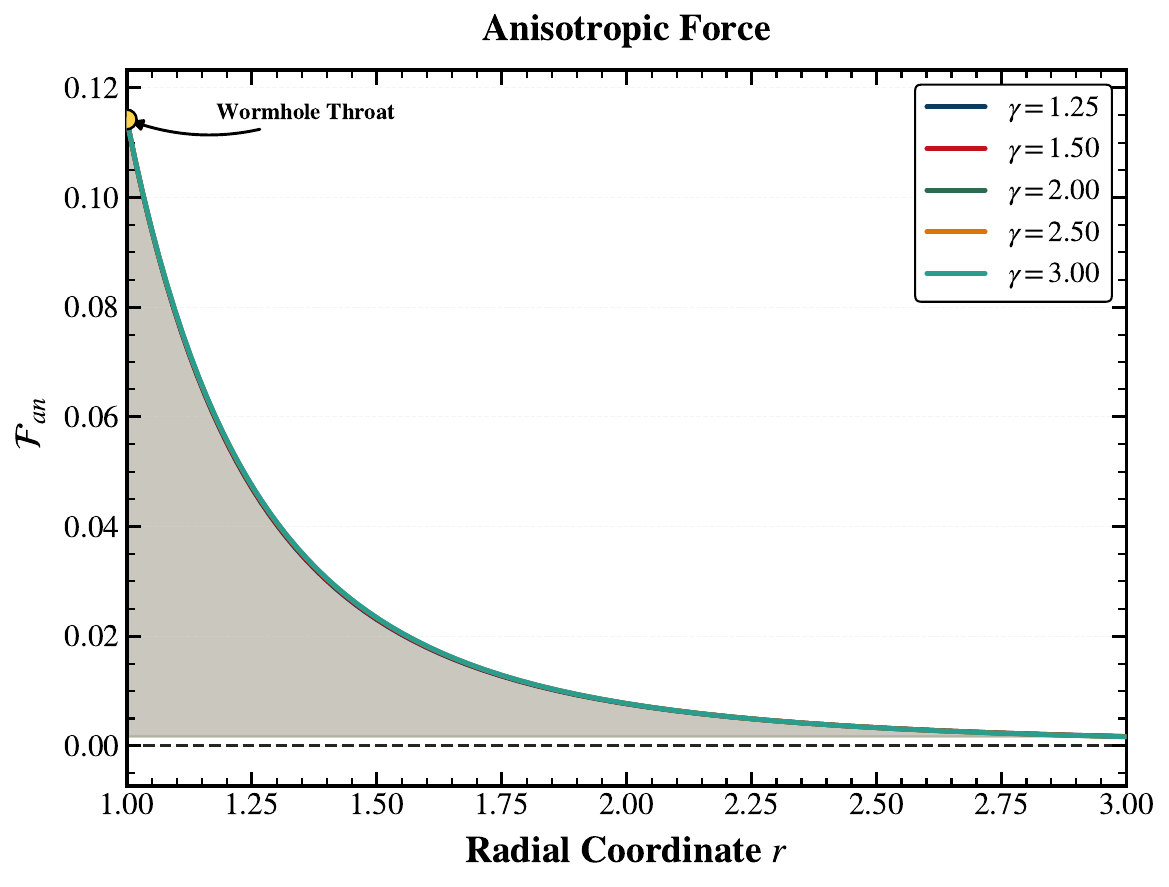}}}
\qquad
\subfloat[]{{\includegraphics[height=2 in, width=3 in]{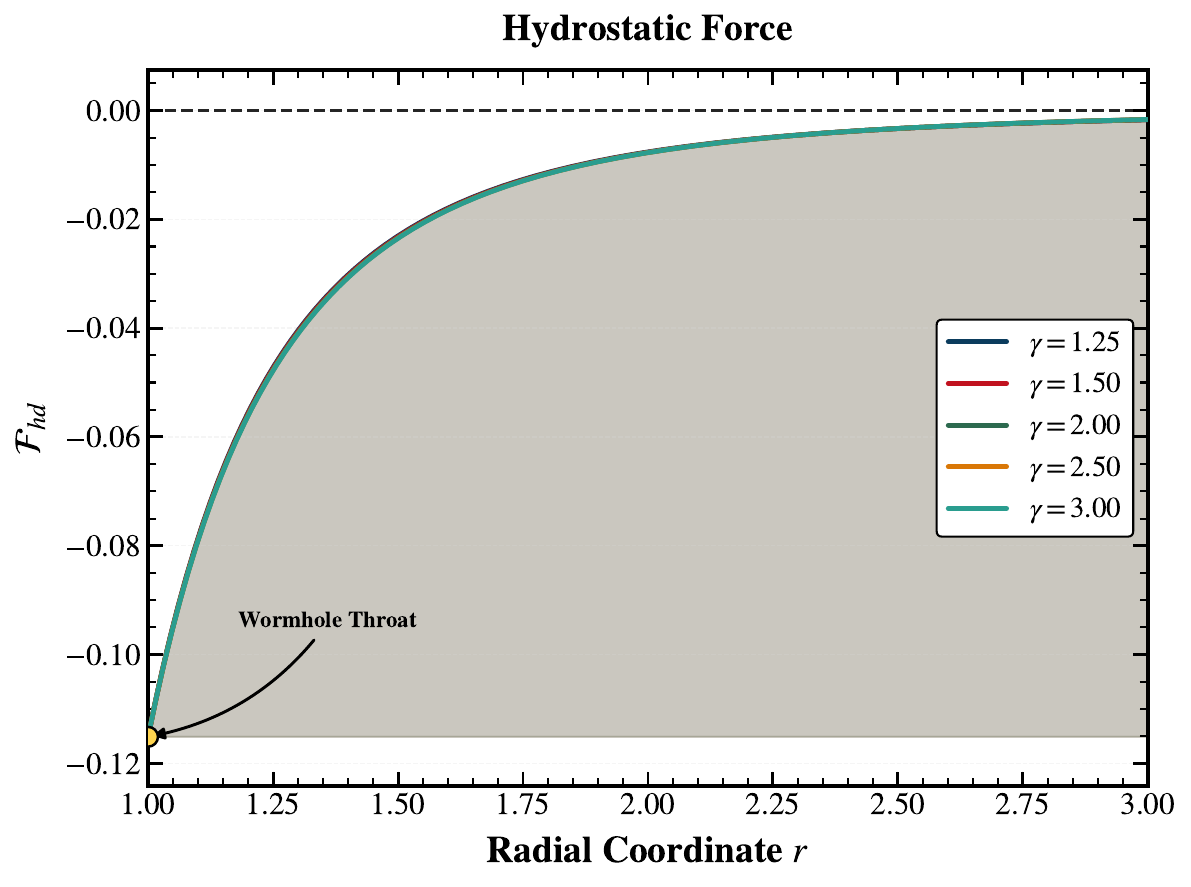}}}
\qquad
\subfloat[]{{\includegraphics[height=2 in, width=3 in]{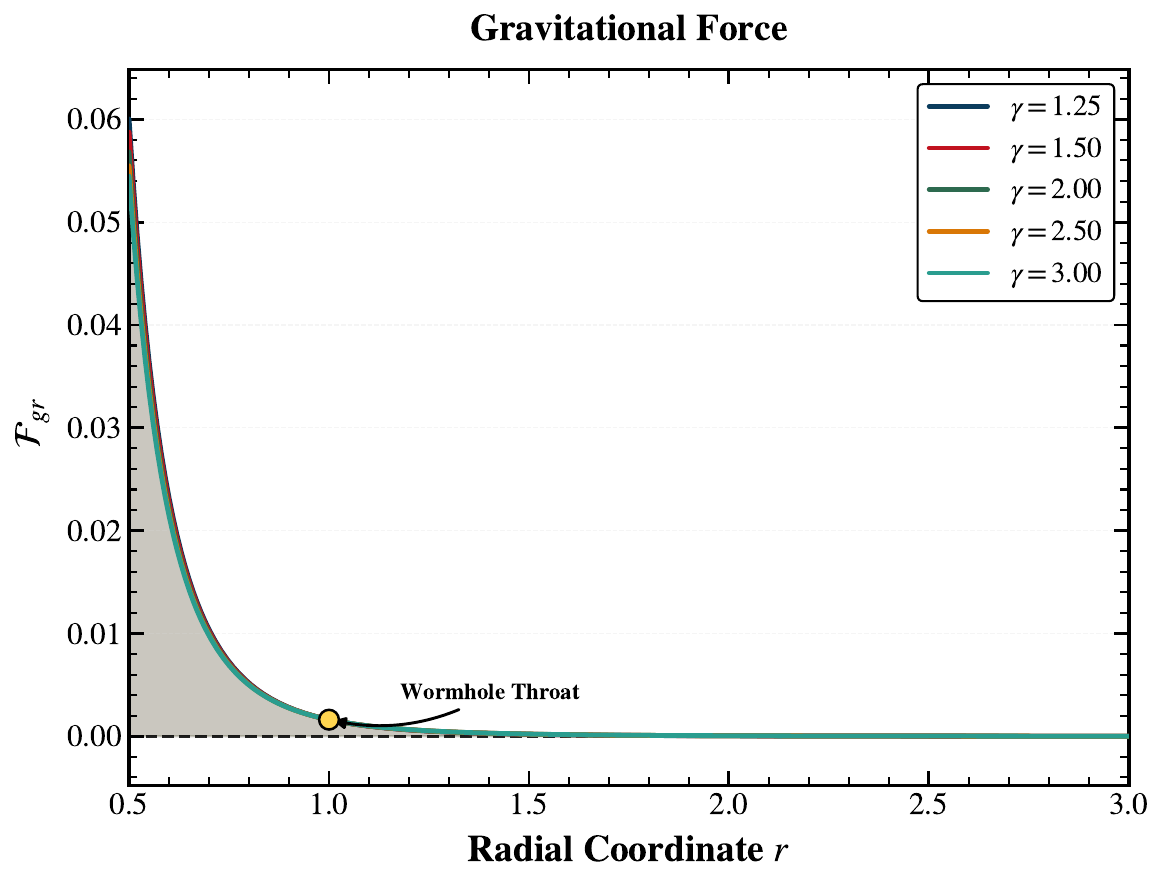}}}
\qquad
\subfloat[]{{\includegraphics[height=2 in, width=3 in]{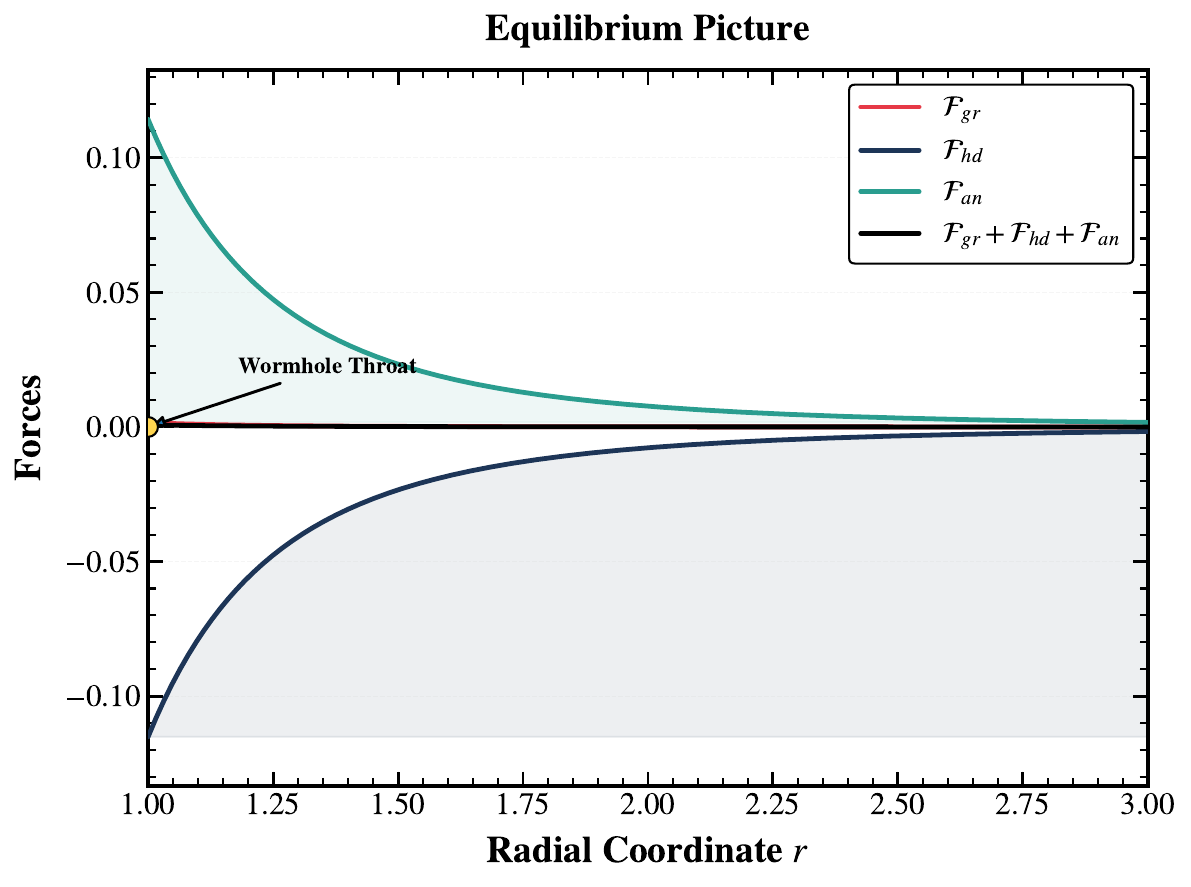}}}
\caption {The plots for profiles of anisotropic, hydrostatic, and gravitational forces, as well as the equilibrium state of the fractional HDE WH for $d=0.05$ and $r_0=1$, for different values of the fractional parameter $\gamma$ have been presented. The equilibrium state of these forces ensures that the wormhole is hydrostatically stable, while the fractional parameter $\gamma$ controls the force magnitude. }\label{10f}
\end{figure}
From the graphical analysis (depicted in Fig. \textbf{\ref{10f}}), it is clear that the inward gravitational force is completely balanced by the hydrostatic and anisotropic forces at any value of the fractional parameter. Therefore, the generalized TOV equation is fulfilled in the whole physical domain.

In other words, the fractional parameter $\gamma$ affects only the magnitude of the forces while preserving the balance between them. Although the increase in the value of $\gamma$ affects the contribution of the hydrostatic and anisotropic forces, the equilibrium holds, which proves that our fractional HDE model produces stable static WH configurations over the considered parameter range.

\section{Non-Singular Core and Structural Complexity}
Apart from the satisfaction of the flare-out condition and energy conditions, the physical feasibility of the traversable WH solution depends on the regularity of the geometry and complexity of the matter distribution. The regularity of the geometry is determined using the Kretschmann scalar of curvature, which is an invariant measure of curvature for spacetimes and can detect curvature singularities. In addition, the complexity factor is considered to estimate the amount of structural complexity that is brought into the system by the anisotropic nature of the matter distribution. The corresponding numerical results are shown in Fig.~11.
\begin{figure}[H]
\centering
\subfloat[]{{\includegraphics[height=2 in, width=3 in]{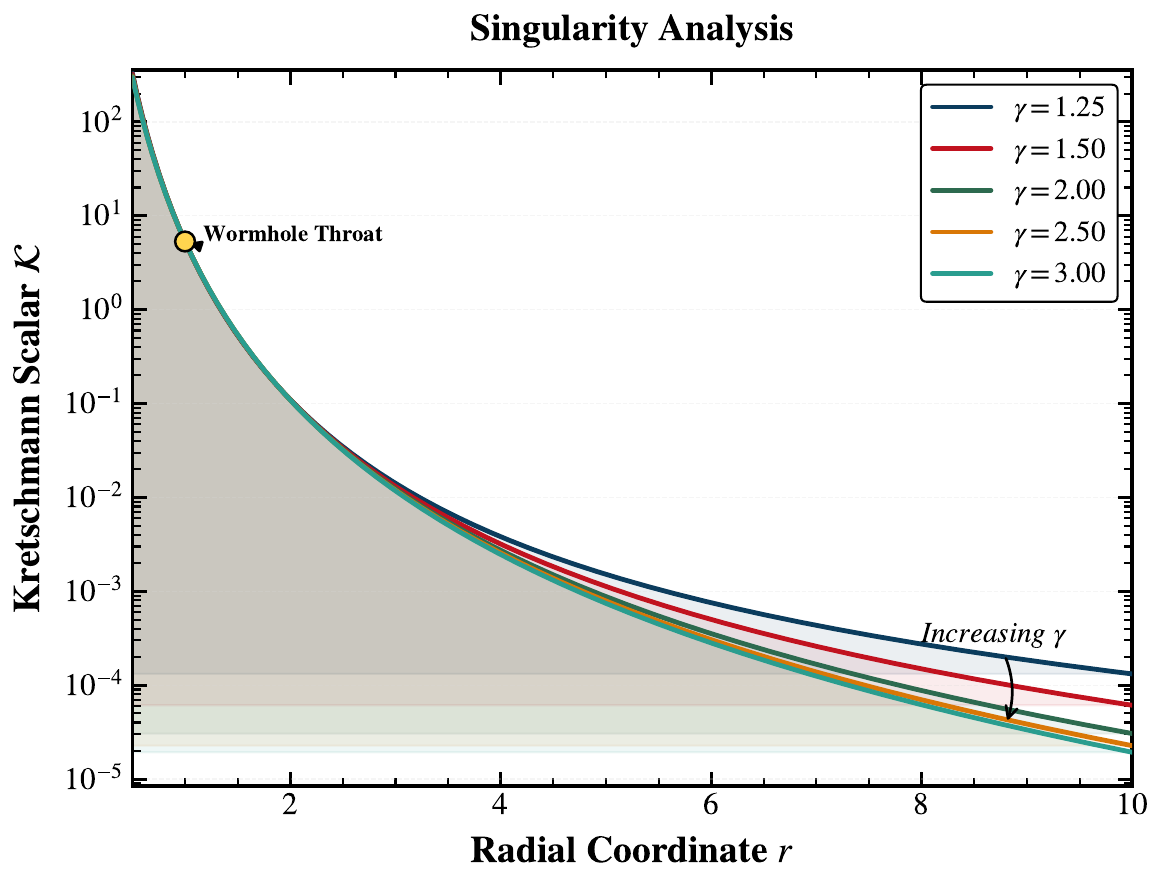}}}
\qquad
\subfloat[]{{\includegraphics[height=2 in, width=3 in]{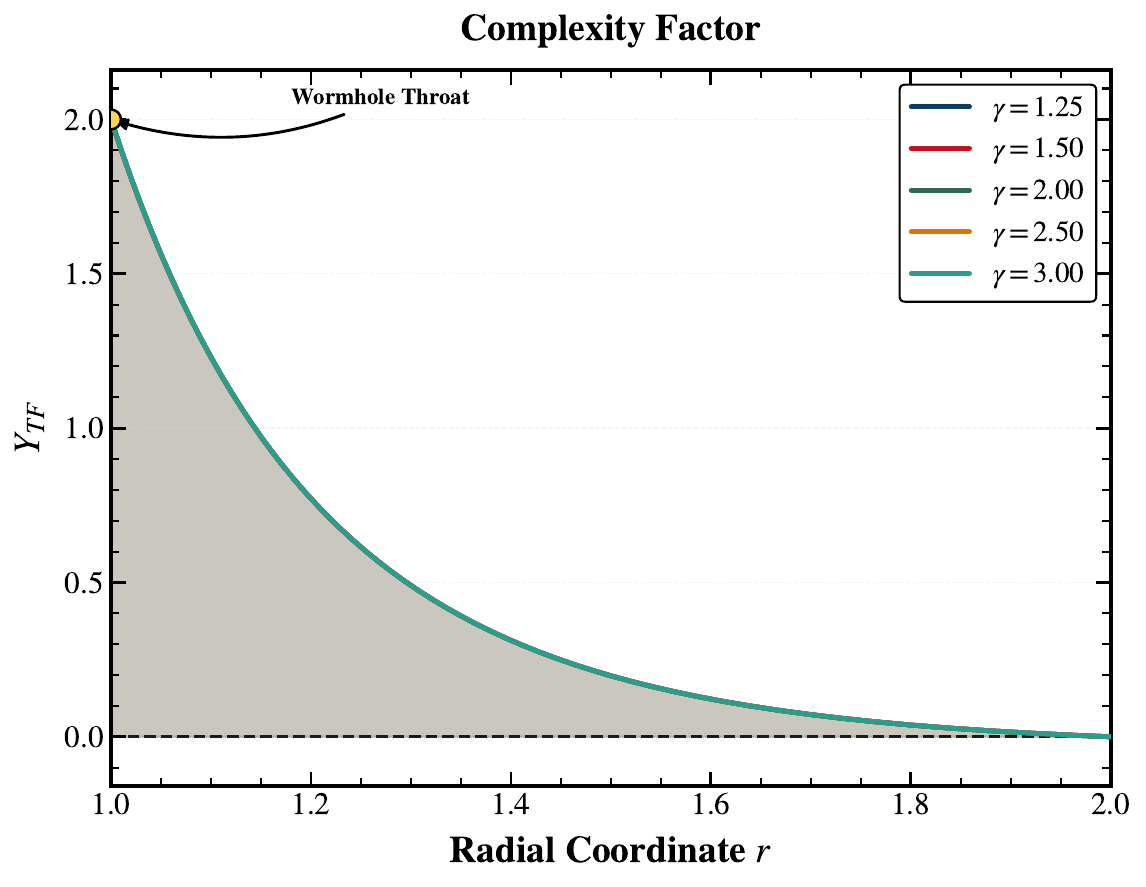}}}
\caption {Graphical representation of the Kretschmann scalar and the complexity factor of the fractional holographic dark energy wormhole for $d=0.05$, $r_0=1$, and for various values of the fractional parameter $\gamma$. The bounded Kretschmann scalar ensures that no curvature singularity exists, and the smoothness of the complexity factor implies that there is an organized structure inside; the fractional parameter $\gamma$ varies the two parameters smoothly without affecting their physical validity.}\label{11f}
\end{figure}
\subsection{Kretschmann Scalar and Singularity Analysis}

To study the presence of curvature singularities, we calculate the Kretschmann scalar,
\begin{equation}
\mathcal{K}=R_{\alpha\beta\zeta\nu}R^{\alpha\beta\zeta\nu},
\end{equation}
a coordinate-invariant measure of the spacetime curvature. Non-divergence of $\mathcal{K}$ within the whole physical region means the absence of curvature singularities in the corresponding geometry \cite{kretschmann1918physikalischen,visser1995lorentzian,morris1988wormholes}.

Figure \textbf{\ref{11f}}(a) shows the dependence of the Kretschmann scalar on the radial coordinate for different fractional parameter $\gamma$. From the figure, it can be seen that the curvature attains its maximal finite value at the throat of the WH ($r=r_{0}=1$) and rapidly decreases as the radial coordinate increases. In all cases, the Kretschmann scalar is positive, smooth, and bounded without any divergence in the whole spacetime region. This implies that the derived fractional HDE WH is curvature singularity-free.

The effect of the fractional parameter is also manifested in Fig. \textbf{\ref{11f}}(a). With an increase in $\gamma$, the Kretschmann scalar decays faster. In other words, the strength of spacetime curvature decreases away from the throat region. Thus, the higher the value of the fractional parameter, the weaker the geometry curvature and the enhanced regularity. These results confirm the existence of a non-singular WH core supported by fractional HDE.

\subsection{Complexity Factor}
The complexity factor serves as an effective tool to quantify the complexity of internal structures of self-gravitating systems through the inclusion of joint contributions of density inhomogeneity and anisotropic stresses \cite{herrera2018new,herrera2018definition}. For the present WH model, the complexity factor can be written as \cite{herrera2019complexity,herrera2019complexitya}
\begin{equation}
Y_{TF}=8\pi(p_{tn}-p_{rd})-\frac{4\pi}{r^{3}}
\int_{0}^{r}\tilde r^{\,3}\mu'(\tilde r)\,d\tilde r.
\label{Complexity}
\end{equation}

The behavior of $Y_{TF}$ has been shown in Fig. \textbf{\ref{11f}}(b). The complexity factor attains its maximum finite value close to the WH throat, while it decreases with the increase of radial coordinate for all the selected fractional parameters. In the end, the complexity factor tends to zero when $r$ increases further. Thus, the influence of density inhomogeneity and anisotropic stresses on the WH becomes weak away from the throat.

Another interesting observation made from Fig. \textbf{\ref{11f}}(b) is that the plots corresponding to different fractional parameters almost coincide in the entire domain of interest. It shows that the complexity factor is not sensitive to the change of fractional parameter and the structural complexity of the WH stays nearly the same. In addition, the positive and finite nature of $Y_{TF}$ indicates that the obtained WH is physically well-behaved and free from any pathology.

Overall, the joint examination of Kretschmann scalar along with the complexity measure reveals that the currently proposed fractional HDE WH possesses a regular non-singular structure along with a finitely complex structure. The above-mentioned features contribute to the validation of the physically acceptable and stable nature of the obtained solutions of the WH.

\section*{Summary}

In the present study, a new class of traversable WH solutions in the framework of GR with fractional HDE has been explored using a variable redshift function. The found shape function ensures the flare-out condition and gives rise to proper two and three-dimensional embedding, which assures the existence of traversable WHs. The fractional parameter $\gamma$ affects the shape function and embedding but does not alter the geometrical nature of the WH solution.

The physical investigation of the spacetime shows that the energy density is positive throughout the spacetime. Even though the radial NEC is violated around the throat and thus exotic matter is available, the tangential NEC and SEC are satisfied. The radial EoS shows phantom behavior, while the tangential EoS is always regular. The anisotropy factor is positive and acts in the outward direction, which prevents the WH from collapsing under the gravitational pull.  Furthermore, the active gravitational mass, compactness function, volume integral quantifier, and equilibrium condition all demonstrate physically acceptable behavior.

Kretschmann scalar shows a finite value throughout the whole space-time manifold, thereby validating the existence of a non-singular core of the WH. The complexity factor also exhibits regularity, which ensures a well-defined internal structure. Thermodynamic parameters such as Hawking temperature, WH temperature, effective pressure, internal energy, work density, energy flux, and specific heat capacity also demonstrate smooth behavior and are devoid of any kind of pathological behavior.

An important result of this study lies in the role of the fractional parameter $\gamma$. With increasing $\gamma$, the geometry of the WH changes, the violations of the radial NEC weaken, the phantom-like nature of the radial EoS is weakened, anisotropic stress is reduced, and smoothness in thermodynamics and curvature is achieved while ensuring traversability conditions. Thus, the fractional parameter acts like a regulator for achieving the desired values and enhances the viability and stability of fractional HDE WHs with reduced exotic matter content.
\section{Conclusion}

In the present work, we have explored a new family of traversable WH solutions sustained by fractional HDE within the scope of GR based on a variable redshift function. The solutions obtained in our study are found to meet all the geometrical criteria for traversable WHs and behave physically and thermodynamically in a consistent manner. From a combined geometrical, physical, and thermodynamical study, it is observed that the suggested model can be used successfully in creating regular WH structures using a smaller amount of exotic matter.

The major result achieved through the present investigation is the key role of the fractional parameter $\gamma$. The graphical analysis shows that the change in the value of $\gamma$ affects both the geometry of the WH and the properties of the matter. As a result of this effect, large values of $\gamma$ lead to a relaxation in violation of the radial NEC, a reduction in the phantom character of the radial pressure, and smoothing of the curvature and thermodynamical profiles. Therefore, the fractional parameter can be considered as an efficient regulator of the physical properties of the WH solution without changing its geometrical properties.

The current work reveals the promise of fractional HDE as an effective model to generate traversable WHs and sets up a strong link between fractional cosmology and WHs. The inclusion of one extra parameter gives rise to new ways to construct physically viable WH solutions that are not only regular but thermodynamically stable as well.

In future studies, researchers can generalize the present study by exploring rotating and charged WHs, different IR cut-offs of fractional HDE models, as well as other gravity theories. It is also interesting to examine the observational features of the solutions obtained in the context of gravitational lensing, quasinormal modes, photon rings, shadow formation, and gravitational wave echoes, as this might be a promising way to test fractional HDE WHs observationally in the future.

\section*{Declaration of competing interest}

The authors have no conflict of interest with respect to the publication of the
present paper.

\section*{Data Availability Statement}

This manuscript has no associated data or the data will not be deposited. [Authors' Comment: This manuscript contains no associated data.]

\section*{ORCID iDs}

M. Rizwan: https://orcid.org/0009-0003-7466-1039\\
Z. Yousaf: https://orcid.org/0000-0001-8227-2621

\section*{CRediT authorship contribution statement}

MR.: Writing – review \& editing, Conceptualization, Writing –
original draft, Visualization, Validation, Software, Methodology, Investigation.
ZY.: Supervision, Validation, Methodology, Formal analysis, Software, Investigation.

\end{document}